# Optical Memory, Switching, and Neuromorphic Functionality in Metal Halide Perovskite Materials and Devices

*Gaurav Vats,[1,2*] Brett Hodges,[1] Andrew J. Ferguson,[1*] Lance Wheeler,[1*] Jeffrey L. Blackburn[1*]*

1. *National Renewable Energy Laboratory, Golden, CO 80401*
2. *Department of Physics and Astronomy, Katholieke Universiteit Leuven, Celestijnenlaan 200D, B-3001 Leuven, Belgium*

**Abstract**

Metal halide perovskite-based materials have emerged over the past few decades as remarkable solution-processable opto-electronic materials with many intriguing properties and potential applications. These emerging materials have recently been considered for their promise in low-energy memory and information processing applications. In particular, their large optical cross-sections, high photoconductance contrast, large carrier diffusion lengths, and mixed electronic/ionic transport mechanisms are attractive for enabling memory elements and neuromorphic devices that are written and/or read in the optical domain. Here, we review recent progress towards memory and neuromorphic functionality in metal halide perovskite materials and devices where photons are used as a critical degree of freedom for switching, memory, and neuromorphic functionality.

## 1. Introduction

Information and communications technologies have the potential to grow to 20% of global energy use by 2030[1] and 50% by 2050. Rising data volumes, filtering, and fast access to the information have led to a paradigm shift from conventional von Neumann architectures to more energy-efficient systems. In this context, brain-inspired systems (replicating the functionality of neurons and synapses), also known as neuromorphics, have emerged as a potential solution. Neuromorphic materials and devices offer opportunities to address impending challenges associated with collecting, processing, storing, and transmitting the vast quantities of complex data predicted for future computing needs. One particularly attractive and relatively underexplored opportunity is in emulating sensory neuronal functions, such as sight or visual perception (Fig. 1). Optically addressable neuromorphic devices and systems have been proposed as alternatives for next-generation visual perception applications, replacing slow and energy-inefficient conventional image capture/processing strategies that are implemented using high-fidelity digital cameras and traditional complementary metal-oxide-semiconductor (CMOS) computing architectures.[2] Low-power environmental (e.g. audio/video/tactile/motion) sensors[2–4] are also a key component of distributed Internet-of-Things (IoT) applications where real-time processing, sensory fusion, and online classification and learning are paramount. These technological needs motivate the research and development of materials and devices with strong inter-coupling between light, charge carriers, and ions that can be incorporated into optically addressable neuromorphic systems. Such systems can encompass both optically stimulated switching between multi-level or analogue-tunable states that can be read out through various electrical means (e.g., conductance, capacitance) and/or states that are read out optically in particular spectral regions (e.g., visible, infrared, gigahertz, terahertz).

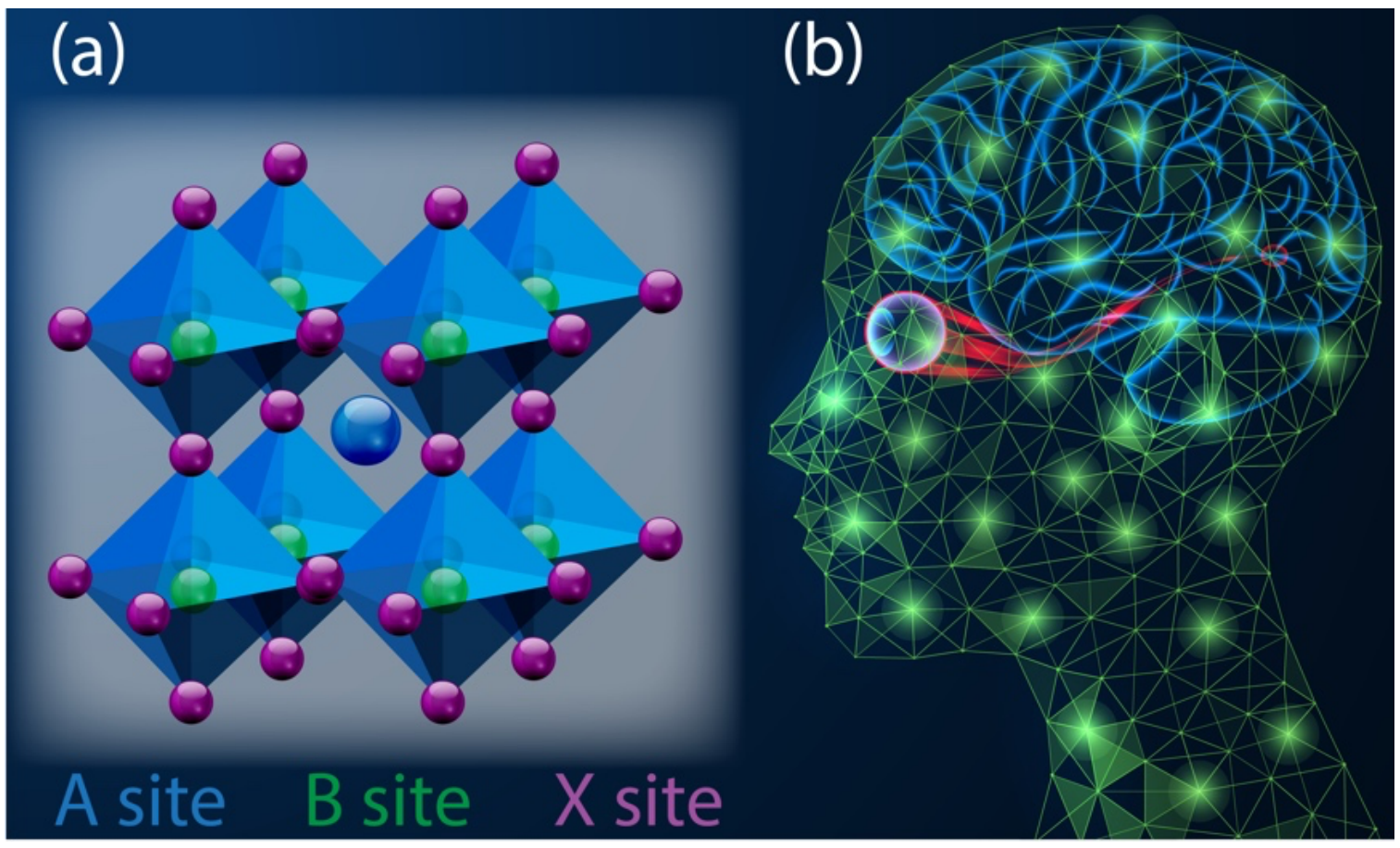

**Figure 1: (a)** The $ABX_3$ perovskite crystal structure that forms the basis for the metal halide semiconductor systems considered in this review for emulation of optical memory, switching, and neuromorphic functionality similar to that exhibited by **(b)** the visual sensory systems and human brain. Image in (b) is licensed under CC BY 2.0, and the color scheme has been modified.

Hybrid metal halide perovskite (MHP) materials have recently made a rapid and dramatic impact on the world of opto-electronics,[5–7] exemplified in large part by their rise to the forefront of next-generation photovoltaic (PV) devices.[8,9] Perovskite refers to the $ABX_3$ crystal structure (Fig. 1a), where A and B are cations and X is an anion and the structure is typically comprised of interconnected $[BX_6]$ octahedra and complementary A-site cations. In MHPs, A is an organic or inorganic cation, B is a metal cation (*e.g.* $Pb^{2+}$, $Sn^{2+}$), and X is a halide ($I^-$, $Br^-$, or $Cl^-$). One key technological advantage of MHP materials is that they can be deposited by simple and cost-effective solution-processing routes. Despite their ease of processing, they also exhibit properties that match or exceed conventional semiconductors (e.g. III-V materials like GaAs) that are relatively expensive to produce.

Beyond their impact in the photovoltaic arena, MHPs also have unique and highly desirable properties for neuromorphic information processing[10–12] that are distinct from the transition metal oxides which have been a dominant focus of neuromorphic research so far. These mixed ionic and electronic conducting materials[13] exhibit large dynamic responses of ions to external stimuli (voltage, temperature, light, strain).[14,15] While relatively facile ion migration in MHPs has been a challenge to overcome for photovoltaics,[16–19] it has only recently been explored as an enabler for neuromorphic functionality.[10,12,20–22] Ionic migration can be used to induce memristive or memcapacitive switching, and if harnessed appropriately, ion dynamics can also encode temporal information in a biologically inspired way to enable faithful artificial synapses – one critical element needed for advanced neuromorphic computing algorithms. Recent studies have demonstrated resistive and capacitive memory in some 3D[20,23] and 2D[21,24] MHP-based artificial synapses, with some systems exhibiting notably small (fJ) switching energies. Importantly, photoexcitation has been shown to dramatically reduce ion diffusion/migration barriers in MHPs.[19,25] Due to the strong optical response and coupling between light-stimulated ionic and electronic transport, photon energy and intensity are powerful tools to manipulate ionic migration activation energies and temporal dynamics, encouraging the exploration of MHPs in optically driven memory and neuromorphic applications.

The 'liquid-like' physical properties of MHPs also offer tremendous opportunities for optical switching in MHP films and devices.[26–28] In this case, low-energy phase transformations, as well as intercalation and deintercalation of a variety of molecular species, initiate reversible structural transformations through the formation of hydrogen, charge-transfer, ionic, van der Waals, and pi-stacked bonds. These transformations are often coupled to large changes in optical absorption that can range from the reversible binary transition between an optically dark and an optically transparent state[28] to a broadly tunable color pallet spanning much of the visible spectrum.[27] In similar fashion to the relatively low ion migration barriers, the low enthalpy of formation for MHP films can enable favorable switching thermodynamics. For example, certain MHPs can be switched between optically transparent and colored states *via* temperature changes as small as 10 °C, in comparison to the 200 °C needed to switch the prototypical phase change material $Ge_2Sb_2Te_5$ that has been explored extensively for optical memory applications.

In this article, we review the progress in the field for utilizing metal-halide perovskites for optical switching, memory, and neuromorphic functionality. We start with a brief discussion of the need for, and potential benefits of, optical read/write neuromorphic materials and devices, followed by a discussion of the beneficial properties and other advantages of MHPs for such applications. We then review the progress made towards the realization of MHP-based optical synapses and related devices such as artificial retinas and vision sensors. Next, we review the literature on structural transformations in MHPs that give rise to large and tunable optoelectronic property (e.g., optical density or emission) switching. Whereas these structural transformations in MHPs have been studied less within the specific context of neuromorphic functionality, there is a great deal of potential in this nascent field of study. We conclude by providing an outlook and perspective towards the ample untapped opportunities for future studies and technologies based on optical neuromorphic functionality in MHP materials and devices.

## 2. Advantages of Optical Strategies for Switching, Memory, and Neuromorphic Functionality

Complex non-linear computations represent an excellent example where neuromorphic systems can excel over their conventional counterparts in terms of speed, efficiency, and accuracy.[29,30] Optical neuromorphic systems may offer further improvements over their electrical counterparts, since optical initiation may be possible using ultrafast pulses/pulse trains that enable faster switching at very low energies per spiking event.[31,32] Photonic synapses, modulated by optical signals, could effectively utilize the unique and promising characteristics of light to yield computing devices and architectures capable of operating with high speeds and bandwidths, with low crosstalk. The non-contact nature of optical initiation/interrogation, along with the ability to tune the optical response of the primary neuromorphic material throughout large portions of the electromagnetic spectrum offers an ample range of opportunities particularly in optical and photonic wireless applications such as light fidelity (Li-Fi). The neuromorphic elements can be optically addressed (written and/or read) and integrated into on-chip platforms, ushering in a new era of fast and energy-efficient applications which are becoming increasingly widespread.

There is also the possibility of integrating optical neuromorphic elements with other next generation computing platforms, such as quantum computing paradigms that rely on spin-photon transduction for information processing, storage, or transmission. Moreover, there is potential for using photonic synapses that exhibit memory functionality to program neuromorphic elements or for real-time image processing and pattern recognition.[33–38] The research in this direction is further supported by emerging solutions to technological bottlenecks for on-chip integrated lasers and light-sources which has motivated the scientific community to consider the benefits of transitioning from fully electronic to opto-electronic and photonic devices.[39–41] Below we identify some of the potential applications enabled by the distinct advantages of neuromorphic data processing strategies based on optical write/read approaches (Figure 2), including examples of potential real-world applications. For a more detailed and broader discussion of photonic and optoelectronic materials and devices for neuromorphic applications, the reader is directed to several recent review articles.[42–44]

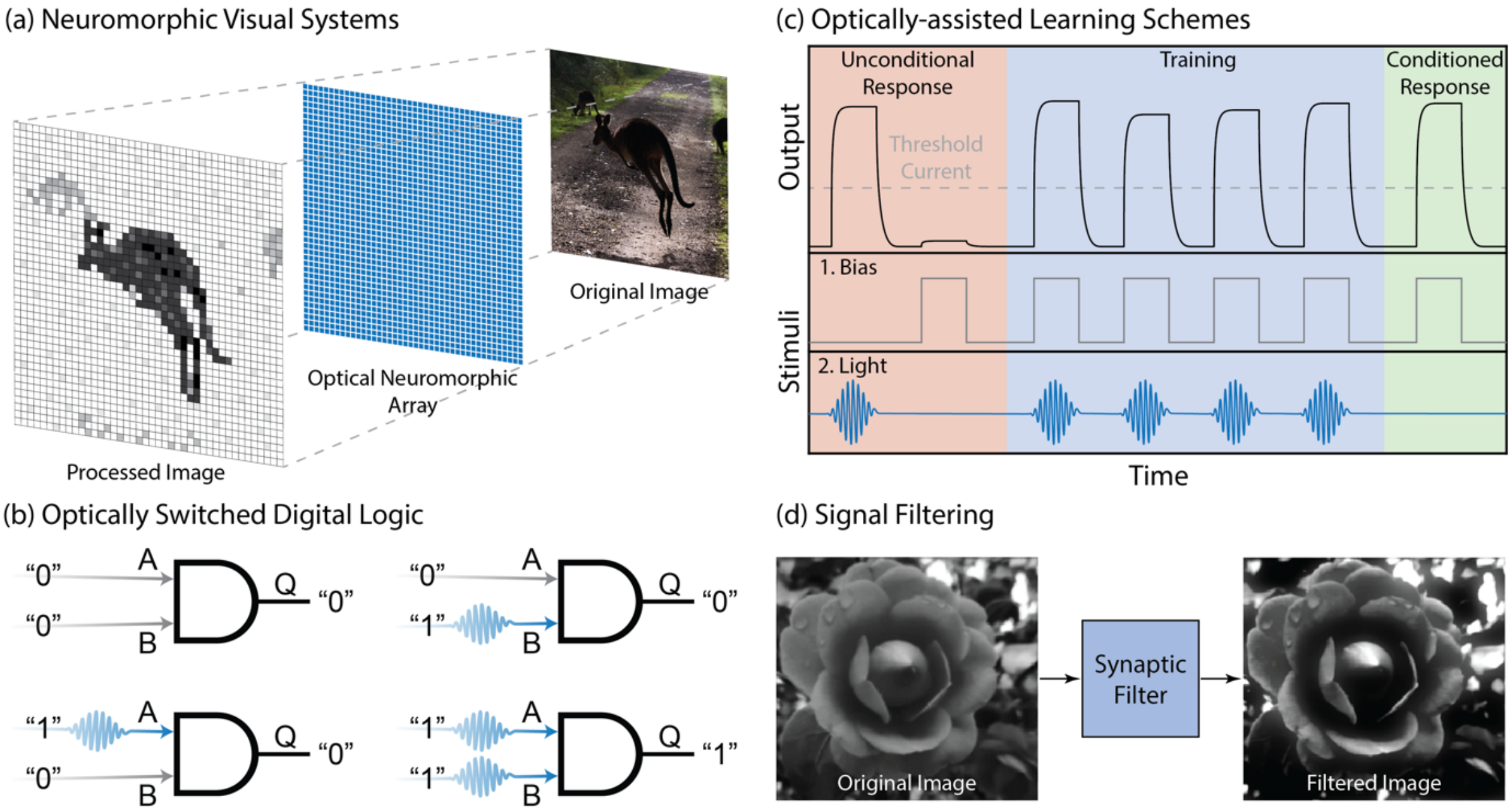

**Figure 2:** Examples of potential applications for optically addressable neuromorphic and synaptic devices and systems. **(a)** Image recognition and processing, such as might be implemented in autonomous vehicles to avoid obstacles. The image "Roo crossing" by ghatamos is licensed under CC BY 2.0 / Original image has been cropped, reflected, and rotated. **(b)** Digital operations, such as those executed by AND logic gates, on optical input pulses. (c) Neural functionality, such as classical or respondent conditioning, where synchronized voltage and optical pulses can be used to train a neuromorphic device to respond to a previously neutral stimulus. (d) Signal filtering, such as the removal of low-frequency noise from an image (adapted with permission from Ref. [35]; Copyright 2020 Elsevier Ltd.).

### *2.1 Neuromorphic Visual Systems*

One of the most obvious applications of neuromorphic visual systems is image or pattern recognition (Fig. 2a), which are often executed by artificial neural networks (ANN) where optical synaptic devices would serve in the synaptic weight component of the ANN. The use of optical or optoelectronic synapses for pattern recognition is currently a growing area of interest in the neuromorphic research community, with much of the effort focused on the identification and development of suitable materials and their integration into the desired sensing and processing architectures.[42–44] Training and evaluation of image/pattern recognition systems is often performed using the Modified National Institute of Standards and Technology (MNIST) database of handwritten digits.

One particularly exciting opportunity is the emulation of biological sight, where real-time visual sensing, data correlation & processing, and dynamic response can be performed more efficiently in an integrated adaptive neuromorphic visual platform. Biological sight, particularly in the case of humans, represents the dominant form of sensory information sourced from the surrounding environment.[45] In biological systems, optical signals are sensed and pre-processed by the eye's retinal circuitry, and the resulting optical information is transmitted by the optic nerves to the visual cortex, where it is processed to allow visual recognition and perception. Therefore, to emulate biological vision, integrated artificial visual systems must be capable of optical detection, image processing, and memorization. In contrast to human vision, which is most sensitive in the visible portion of the electromagnetic spectrum, artificial systems could be capable of detecting and processing optical information across a much broader wavelength range, enabling neuromorphic visual systems capable of sensing their environment at wavelengths optimal for the specific application (e.g., night vision using infrared detection). The ideal artificial neuromorphic visual system will have tunable spectral response (e.g., broadband or narrow band) and high photosensitivity with the ability to adapt to the intensity of the incoming optical signal.

A key inspiration for optical neuromorphic devices and architectures relates to real-time spatio-temporal image processing (e.g., Fig. 2a), and an exemplary system that provides inspiration is an *all-analog artificial retina system*. These biologically realistic light sensors can directly transduce incident light with different intensities (brightness) and wavelengths (color) into analog conductance modulation for real-time adaptive learning in applications that require the ability to 'see' and react to surroundings, such as internet-of-things sensors and self-driving (or autonomous) vehicles. Existing conventional visual detection and perception systems are built on complex integrated components that capture images using high-fidelity digital cameras and process them with algorithms implemented by separate CMOS-based memory and processing hardware, resulting in high energy/power consumption. To address this issue, research has begun focusing on neuromorphic visual systems built on the foundation of optoelectronic synaptic devices.[42–44,46]

As one specific example of the energy benefits of neuromorphic visual systems, significant energy savings are predicted for autonomous vehicles when all dynamic driving functions are fully automated (Level 5).[47,48] However, this requires that the various sensing, processing, and response functions are performed more efficiently than by conventional power-hungry data processing systems. It has been proposed that optical synapses can be paired with e.g. spiking neural networks for all-analog visual perception and processing. Such an architecture with enhanced real-time visual sensing and processing offers opportunities for significant improvements in the driving performance (accuracy and efficiency) over sometimes fallible human drivers. This will help to enable complete electrification and full autonomy in the transportation sector, yielding dramatic reductions in power/fuel consumption and improved vehicle mileage to reduce costs and carbon footprints associated with the movement of people and goods. Such goals can only be achieved via the design and development of tailored optical neuromorphic materials and devices.

### *2.2 Logic Functions*

Modern computation is still based on the von Neumann computing architecture and is built on the foundation of digital bitwise logic and arithmetic operations.[49,50] Optically addressed synaptic logic devices for neural network applications integrated with complimentary-metal-oxide-semiconductor (CMOS) processing have been considered for more than three decades.[51] More recently, some focus has been devoted to the execution of such operations using neuromorphic elements and systems to overcome the issues associated with physical separation of, and the requirement to transfer information between, the memory and processing components (i.e., the von Neumann bottleneck and memory wall problem).[52,53] There are also demonstrations of several logic operations using optoelectronic synaptic devices (Fig. 2b),[36,54–56] including fabrication of the universal logic gates (NAND and NOR) that can be used in networks to realize every possible logic gate.

*2.3 Emulation of Neural Activities*

Memristors represent devices that emulate synaptic weighting by performing in-memory computation on an external stimulus since the device memory is held in the tunable physical state (e.g., conductance). This property enables imitation of biological sensory memory (SM), short-term memory (STM) and long-term memory (LTM) processes as well as a number of synaptic behaviors: excitatory postsynaptic current (EPSC), inhibitory postsynaptic current (IPSC), paired-pulse facilitation (PPF), spike-voltage dependent plasticity (SVDP), STDP, spike-number dependent plasticity (SNDP), spike-frequency dependent plasticity (SFDP), and transition from SM to STM. An overview of various synaptic functionalities could be found in the references.[30,57] Although most examples of memory functionality (learning and forgetting) have been demonstrated using electrically-stimulated devices, an interesting prospect is the realization of optical neuromorphic memory and neural emulation (Fig. 2c). Several studies have shown that both memory and forgetting can be stimulated by pulse trains, with the efficacy (memory level and retention time) correlated with the intensity, duration, and quantity of the pulses.[58–61] However, optimal learning and information processing in biological systems relies on more complex neural mechanisms that, if implemented in artificial systems, also promise to meet the demands for energy-efficient deep learning and advanced neuromorphic functionality. These mechanisms include multi-factor learning, dendritic computation, and composite plasticity.

There are several examples of exploiting the nature of the input stimuli to train the neuromorphic element/device via associative or aversion learning protocols. For instance, the output current of a 2D transition metal dichalcogenide transistor device could initially only be increased above a threshold using optical pulses, whereas electrical pulses did not induce the same conductance enhancement.[36] After training the device with synchronized optical and electrical pulses, the device was conditioned to enable switching above the conductance threshold using purely electrical pulses, simulating classical or respondent conditioning. Aversion learning has also been mimicked using a combination of voltage and optical pulses in a silicon nanocrystal (Si NC) phototransistor.[56] Voltage pulses alone induce a current oscillation and a slight increase in the device conductance, but synchronized voltage and optical pulses result in a gradual reduction in the device conductance, simulating aversion.

Emulation of biological synaptic plasticity was subsequently facilitated by photogating of a transistor due to the heterojunction between an organo-metal halide perovskite and a Si NC layer in the transistor channel.[37] This synaptic plasticity enabled learning of a biased and correlated random-walk (BCRW), although full hardware implementation would require fabrication of an artificial neural network of perovskite-enhanced Si NC synaptic devices interconnected with neuronal devices.

*2.4 Signal Filtering*

Another synaptic function in biological systems is spike rate dependent plasticity (SRDP), which allows realization of signal filtering functionality (Fig. 2d) and will be an important component of artificial neural networks. Both electrically and optically stimulated devices have displayed SRDP, and there have even been demonstrations of signal processing based on the frequency properties of the input. For example, the frequency of optical illumination of an electrochemical $MoS_2$ field-effect transistor was used to demonstrate both potentiation, or enhancement of the synaptic weight (at frequencies exceeding ca. 10 Hz) and depression, or reduction of the synaptic weight (at frequencies less than ca. 10 Hz).[62] Both low-pass and high-pass optical filters have been demonstrated, using transistors with channels composed of printed semiconducting single-walled carbon nanotubes[63] and spin-coated conjugated polymer/inorganic halide perovskite quantum dot blends,[64] respectively. More recently, an optoelectronic device with an active layer layer comprising a blend of Si NCs in methylammonium lead iodide (MAPI) demonstrated high-pass signal filtering, enabling image enhancement via the removal of low-frequency noise below the cutoff frequency of 4.8 Hz (Fig. 2d).[35] Despite these impressive demonstrations, which highlight the potential for effective information processing and transmission, the operating frequencies of the devices remain quite low. This observation points to the need for further investigation of the ultimate limits for the frequency-dependent signal filtering and the potential applications the performance characteristics may enable.

## 3. Advantages of MHPs for Optical Switching, Memory, and Neuromorphic Devices

A crucial property of MHPs that has helped drive their research and integration in photovoltaics is a large optical absorption coefficient and steep optical absorption onset. As an example, the prototypical MHP $CH_3NH_3PbI_3$ (often abbreviated $MAPbI_3$ or MAPI) has an optical absorption coefficient of > 5 x $10^4$ $cm^{-1}$ at energies just above the optical bandgap (~1.57 eV or 789 nm) and >1 x $10^5$ $cm^{-1}$ above ~2.1 eV (590 nm).[65] This strong absorption in the visible region of the solar spectrum enables solar cells to be fabricated from very thin layers of MHPs, i.e. < 500 nm. The absorption spectrum can be broadly tuned in MHPs through several parameters. Changing the halide identity (X site) has a large impact on bandgap, which increases from iodide to bromide to chloride. The impact of metal identity (B site) has a smaller, but non-negligible impact, with e.g. Sn compounds having smaller bandgaps than Pb compounds. The organic (A site) orbitals do not typically contribute significantly to band-edge states and optical transitions, but can do so in cases with hybridized (i.e. charge transfer) orbital character.

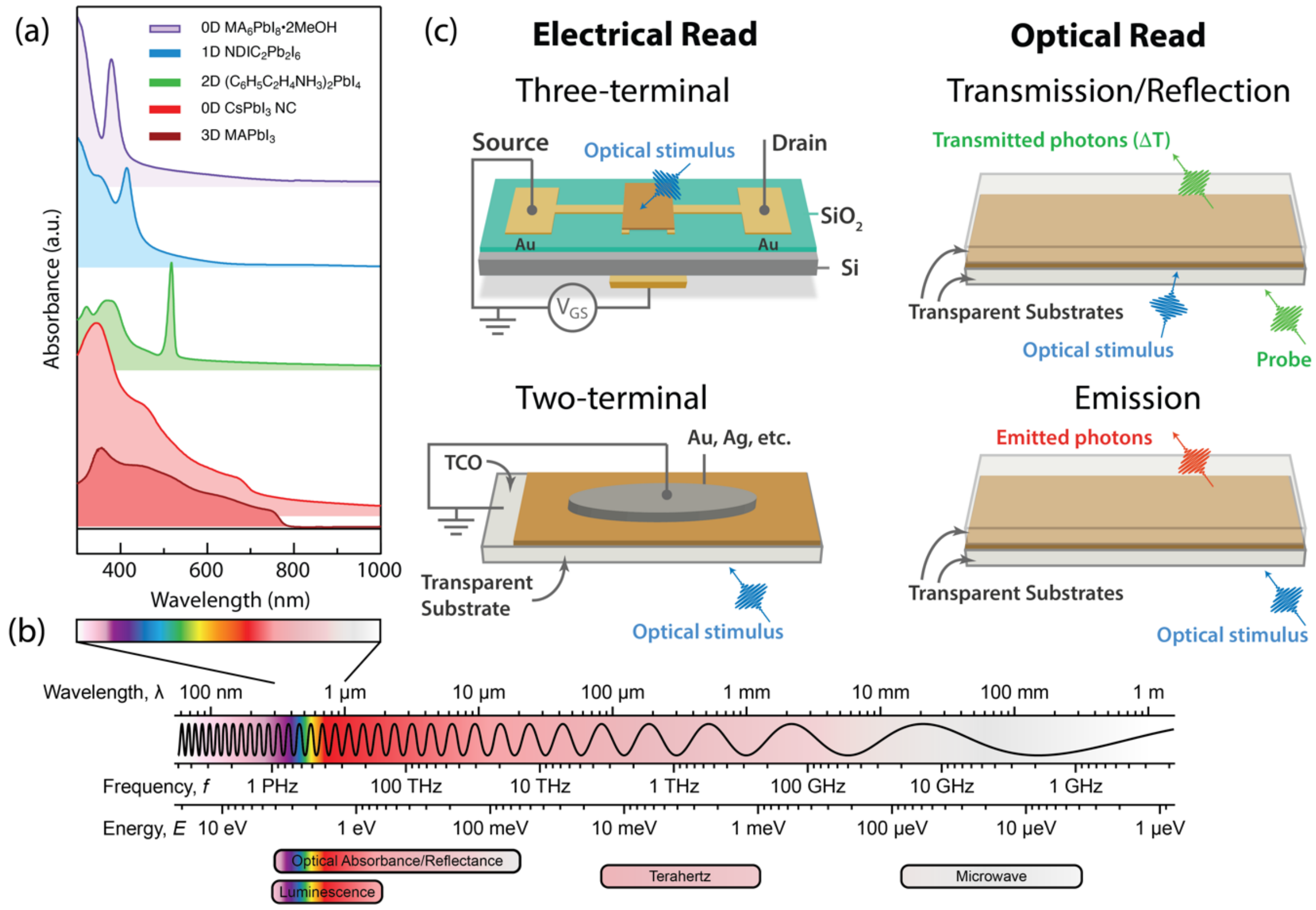

**Figure 3: (a)** Modulation of iodide-based metal halide semiconductor absorbance spectra by A-site cation-driven manipulation of the lead iodide octahedral dimensionality and connectivity, along with one example (0D $CsPbI_3$ NC) where the absorbance spectrum is tuned by quantum confinement at the material level (0D nanocrystals or quantum dots in this case). (MA = methylamine). **(b)** Schematic of the electromagnetic spectrum, along with various regions of the spectrum where optical neuromorphic/memory elements can either be stimulated or probed by optical means. **(c)** Schematics of optically stimulated elements that are either read electrically (left) or optically (right). For 'optical read' elements, the stimulus can be optical, electrical, or thermal, but only optical stimulus is depicted for simplicity. Electrically stimulated 'optical read' devices would require transparent substrates with e.g. transparent conducting oxides (TCO).

MHP dimensionality is another powerful structural knob that can tune optical, electrical, ionic, and magnetic properties (Fig. 3a). Low-dimensional MHPs can be realized by confining *material* dimensions *or* by adjusting stoichiometries to achieve 2D, 1D, or 0D arrangement of the lead halide framework. Depending on the material dimensionality (*vide infra*), temperature, and photon fluence, photoexcitation of an MHP can either produce uncorrelated (free) electrons and holes or Coulomb-bound electron-hole pairs (excitons). In most bulk (3D) MHPs, dielectric screening leads to exciton binding energies that tend to be less than thermal energy (kT) at room temperature, and the dominant photogenerated species are free charge carriers. Larger exciton binding energies in low-dimensional MHPs lead to tunable absorption onset that is often characterized by a sharp optical transition (Fig. 3a). Excitons in these systems are often stable at room temperature and must be dissociated to produce free charge carriers. In the context of neuromorphic functionality, the strong and highly tunable photon absorption in MHPs bodes well for optically modulating and/or reading out synaptic weights in various regions of the electromagnetic spectrum (Fig. 3b and 3c). Electrical read schemes are the most prevalent in the literature. However, as switching energies are reduced, optical read schemes become more attractive to reduce confounding electrical heating effects (Joule heating) that may affect the state of the device. In contrast, larger volumes of material are typically needed to ensure stable readouts in the optical domain instead of the nanoscopic filaments that can be achieved with electrical current switching. The extraordinary absorption coefficients of MHP materials yields more flexibility toward miniaturization and higher density optical read memory devices.

Charge carriers (electrons and holes) have sizeable mobilities in MHPs. Charge carrier mobilities in 'dark' (i.e., not photoexcited) MHPs tend to be low to moderate. The gate field in field-effect transistors (FETs) can be screened by ion migration and room temperature FET mobilities for $MAPbI_3$ are often in the range of $10^{-5} – 10^{-4}$ $cm^2V^{-1}s^{-1}$,[66] although modifications of thin-film morphology and source-drain contacts can increase these mobilities up to ca. 0.5 $cm^2V^{-1}s^{-1}$.[67] Tin compounds tend to have relatively high mobilities compared to their lead counterparts, with early studies finding room temperature Hall mobility of ca. 50 $cm^2V^{-1}s^{-1}$ for pressed pellets of 3D $CH_3NH_3SnI_3$[68] and FET mobilities of ca. 0.1 – 0.6 $cm^2V^{-1}s^{-1}$ for spin-coated 2D $(C_6H_5C_2H_4NH_3)_2SnI_4$ that take advantage of good in-plane transport in the layered compounds.[69] Measured carrier mobilities are typically much larger in the excited state of photoexcited MHPs (relative to dark mobilities), but the reported values vary widely according to the particular measurement technique and significant debate still exists in many cases for the attribution of measured values to either the electron, hole, or both.[70] High carrier mobilities and long carrier lifetimes instill many MHPs with exceptionally long carrier diffusion lengths that can exceed 1 μm. With regards to optical switching and memory, the low/high carrier mobilities for dark/photoexcited MHPs is advantageous since this discrepancy can provide low dark current and large optically induced conductance switching ratios.

Charge carrier transport is often inextricably linked with ion migration in MHPs. Many studies have shown intrinsic and extrinsic ion migration in MHPs under a variety of applied fields.[71–73] Both A-site cations and X-site halide anions can migrate, especially under applied electric fields (Fig. 4a) with halide anion mobility being most pertinent near room temperature. The activation energy, $E_A$, is substantially lower for halide migration on surfaces and grain boundaries, relative to bulk migration, *and* decreases dramatically under light illumination (Fig. 4b).[14] A, B and X site variations enable tuning of the metal halide bond energy that governs the halide vacancy concentration, thus halide motion, and add another knob for tuning $E_A$. Many extrinsic species can also diffuse or migrate into MHPs, including metal atoms from electrodes, dopants from adjacent electron or hole transport layers, and vapor-phase species from the environment (*vide infra*). Optically stimulated ion migration in MHPs has been exploited to realize novel functionality and devices, including switchable PV windows,[26] electronic ratchets,[74] and optical synapses.[75]

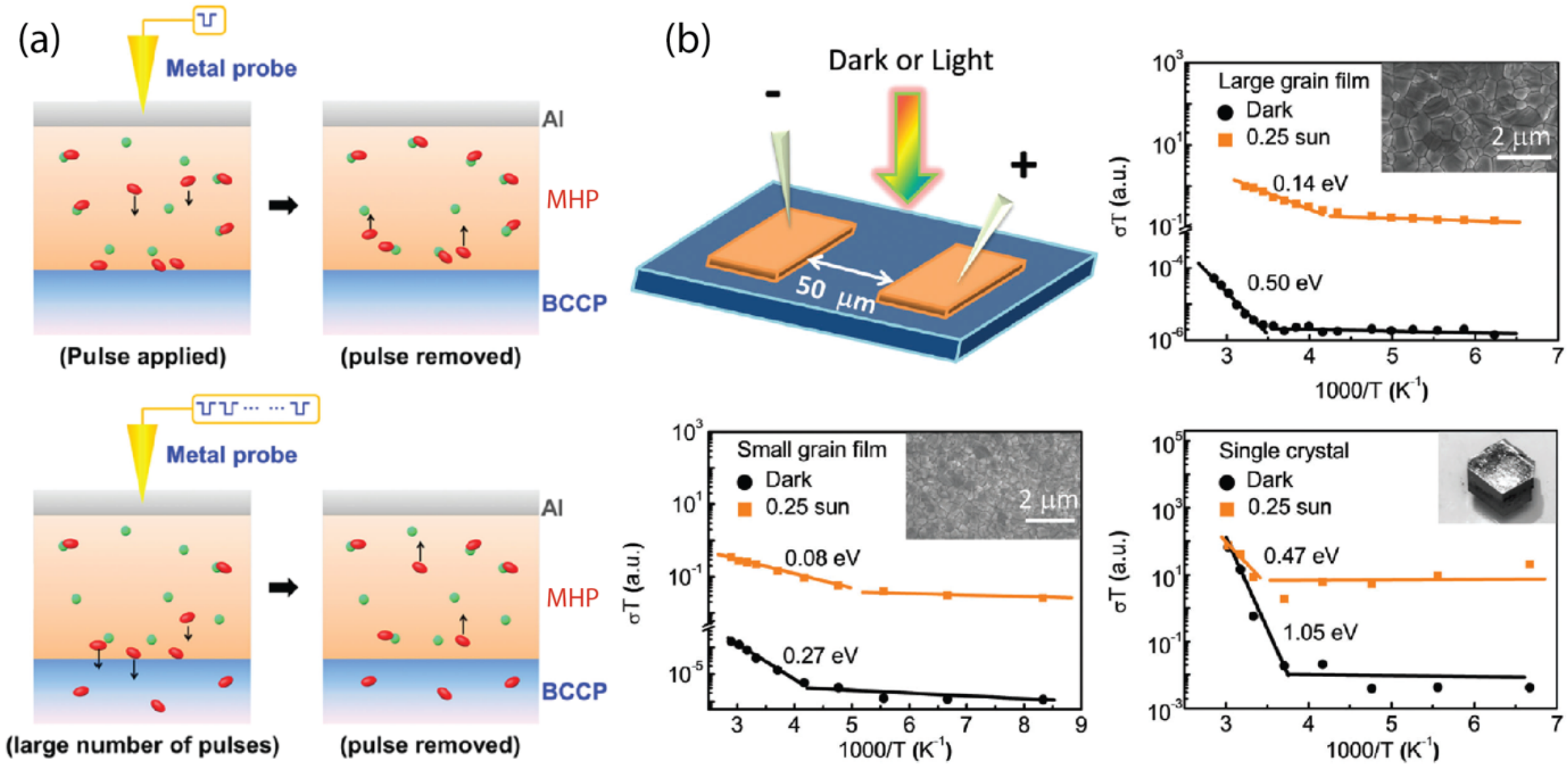

**Figure 4:** (a) Schematic showing ion migration due to applied pulse and ion migration as well as diffusion (and trapping) on applying multiple pulses followed by a return of the charge carriers (anions, red color) and vacancies (shown in green)) in the bounded states on the removal of the applied pulses. Adapted with permission from reference [76]. (b) Measurement schematic and experimental results for temperature-dependent ionic conduction in large-grain film, small-grain film, and single crystal of $MAPbI_3$, either in the dark or under 0.25 sun illumination. Reproduced with permission from reference [77].

Finally, MHP materials have near-zero formation energies and exhibit polymorphism at temperatures that resemble organic materials more than inorganic ones.[78] The low-energy structural properties enable a variety of ways to control large changes in optical density and tunable light emission, including: ion motion, ion exchange, crystal transformation, crystal precipitation, and a variety of intercalation chemistries. The structural transformations are often reversible and can be controlled using external stimuli such as light, heat, or electric field. These low-energy transformations are the bane of photovoltaics researchers,[79] who seek to maintain materials properties over long lifetime and in harsh conditions. However, there is building excitement around leveraging the transformations for new technologies,[80] including thermochromic windows[81–83] and switchable photovoltaic devices.[26,84] Such transformations are also promising, but largely unexplored, for optical memory or neuromorphic functionality.

Information about neuromorphic synaptic devices could also be found in the recent related reviews[11,61,85,86] and references therein. Though there have been several reports on synaptic devices based on MHPs operated by electrical and light-assisted devices, the research is still in infancy and far from deployment for real-life applications. One of the major roadblocks to their deployment is the environmental and chemical stability of several well know MHPs. This stability issue can be tackled by using moisture-resistant layers, encapsulation, and by chemical passivation routes. A detailed discussion of this can be found in the following articles.[87–89] Other important parameters that require attention are low endurance and retention times and scalability (restricted to optical diffraction limit) to compete against existing CMOS platforms.

## 4. Mechanisms for Optical Switching and Memory in Metal Halide Perovskites

The broad range of morphologies, heterostructures, and switching mechanisms, together with tunability-via-light, suggests perovskite-based systems as ideal neuromorphic candidates with multiple intrinsic state variables and rich dynamics similar to biological systems. Based on the incoming stimuli and resulting outputs (e.g., current, optical changes, etc.), we consider two primary categories of optical write/read MHP synaptic and memory devices for neuromorphic applications: (a) light-stimulated and electrically read and (b) optically read devices (Fig. 3c). In terms of light-triggered *initiation* of synaptic elements, above-bandgap photons can generally produce a photoconductance response,[90] especially in weakly screened semiconductors without large excitonic effects (but also in many cases within excitonic semiconductors[91]), that can be modulated by a number of mechanisms. Light energy can also induce structural changes,[92,93] lattice relaxation,[94] diffusion of elements/ions and vacancies[95] and even reversible chemical reaction(s).[96] In terms of light-mediated *read-out* of synaptic elements, the two primary handles are the material's optical complex refractive index or dielectric constant (which can be detected *via* how light is absorbed, transmitted, and/or reflected by the medium) and/or photons emitted by the material, both of which can be modulated by various stimuli through several mechanisms. Here we briefly discuss several of the observed and hypothesized mechanisms underlying optical switching, memory, and neuromorphic behavior in MHP-based systems before reviewing progress towards each of the two classes of materials/devices in Sections 5 and 6.

### *4.1 Resistive Switching*

Resistive switching mechanisms in a material may include thermochemical reactions, ion migration, charge trapping and de-trapping, interfacial reactions, charge transfer, electrochemical redox reactions, and/or conformation change.[97] The functionality and operational regimes of a synapse depend on the resistive switching mechanism(s) and the impact of the controlling stimulus (e.g., applied electric pulse or light-illumination) on material/interfacial structure and properties. Dielectric constant (responsible for resistive states) and charge transport (the driving factor for conductance output signals) play vital roles in the context of optically and electrically triggered devices. Dielectric constant depends on surface charges, ionic displacement, dipolar field and electronic contributions. Charge transport characteristics are governed by the distribution of the electronic cloud and chemical bonding inside the materials,[98,99] which impact mobility, lifetime, diffusion lengths, and carrier effective masses.[98] MHP resistive switching devices working on both electronic and photonic inputs can be impacted by surface/interface dipole effects, interfacial band offsets, ionic displacements/migration, and charge trapping/de-trapping. The impact of any mechanism can be controlled, in part, by engineering the device geometry, with devices typically sub-categorized as 2-terminal and 3-terminal devices (Fig. 3c, and further discussed in Section 5), or by the material composition, dimensionality, and quality (further discussed in Section 6).

#### *4.1.1 Charge Trapping*

A prominent resistive switching mechanism is charge trapping, which can occur due to defects induced by local structural distortions and dangling bonds,[21,100] defects across semiconductor/dielectric or MHP/polymer interface,[100] and engineered potential wells across a heterojunction.[64] In electronically driven systems trapping and de-trapping are controlled by applying and removing electric fields. Initially, ions or vacancy defects are trapped due to the applied electric field followed by reversal on removing or flipping the direction of the applied field with some time delay. In photo-driven devices, light energy is used to trigger trapping of photoinduced charges[100] and/or drift and diffusion of ions, often in response to the photoinduced field produced by trapped charges,[59] followed by electric de-trapping.

*4.1.2 Ion Migration*

Ion migration, in response to an external stimulus (such as an applied electric field, temperature change or light illumination), is a prominent strategy for controlling the output of a synaptic device.[24,101–103] In some cases, ion migration can be tuned by an additional intermediate semiconducting layer (Fig. 4a),[20,76,104] and ion-migration can also lead to the formation of conductive filaments.[105,106] The amplitude of the applied stimulus energy (electric potential, heat transfer or charge injection) governs the extent of ion migration and thickness of the conductive filament.[96] The migration could be reversible (fully or partially) or irreversible depending on the trade-off between the material's phase equilibrium and the applied stimulus energy. This range of reversibility can facilitate short-term plasticity (STP) or long-term plasticity (LTP), as well as transitions between STP/LTP as a function of stimulus frequency, energy, or duration. Ion migration is a well-accepted mechanism in MHPs and light illumination has been shown to enhance ion migration.[14][74] In a seminal study, Xing et al. demonstrated that illumination dramatically decreased ion migration activation energies in lateral two-terminal $CH_3NH_3PbI_3$ (also known as $MAPbI_3$) devices (Fig. 4b).[14] Activation energies for halide migration were found to decrease with decreasing grain size as well.

*4.2 Structural Reconfigurations*

There is in an array of mechanisms to *structurally* reconfigure perovskite materials to yield dramatic swings in optical density or emission properties that could be leveraged in optical memory and neuromorphic devices.[80] The mechanism of transformation requires significantly less energy compared to conventional semiconductor materials. Dynamic switching of optical properties is due to changes in the connectivity of the metal halide octahedra network. Extended 3D $AMX_3$ perovskite materials are isotropic with corner-sharing octahedra (Fig. 5a), where octahedral metal (typically Pb or Sn) centers share halide ligands with adjacent metal centers. The materials exhibit a continuous band structure. The bandgap of the extended 3D perovskite spans <1.2 eV for Sn-Pb iodide alloys (cite) up to > 3 eV in lead chloride analogs.[107]

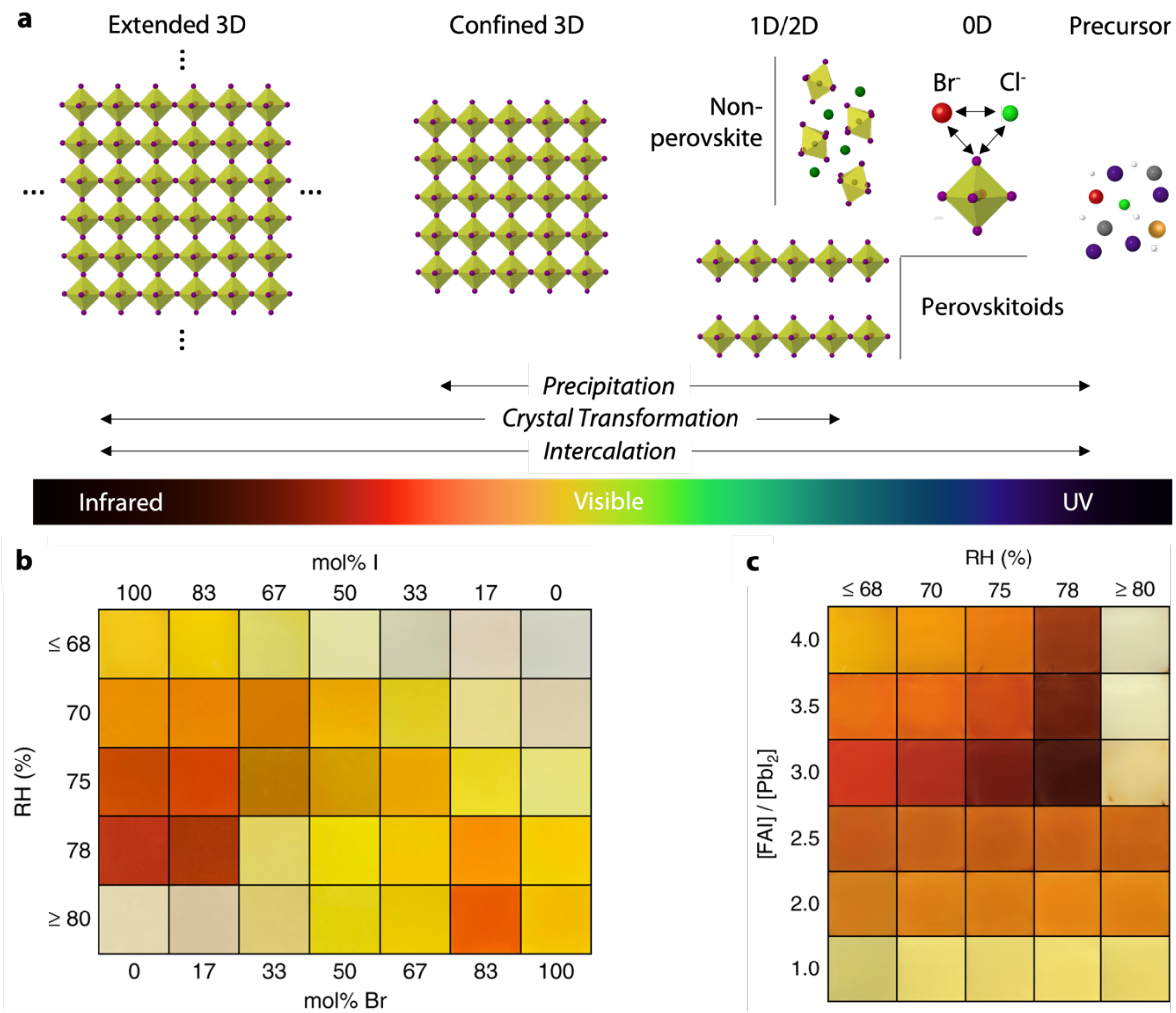

**Figure 5:** (a) Overview of structural transformations that change the connectivity of the metal halide octahedra network and three different mechanisms of transformation and approximate dynamic range for a single halide family of materials, as described in Section 6. (b,c) Photographs of room-temperature reversible transformation of $FA_{n+1}Pb_nX_{3n+1}$ (FA = formamidinium, X = I, Br; $n$ = number of layers = 1,2,3… ∞) films in response to humidity or alcohol environment, where the ratio of I to Br is varied (b) and the ratio of film precursors ($FAI/PbI_2$) is varied (c). Adapted from [108]

Within a halide family, the characteristic bandgap can be increased and transitioned from band-like to discrete molecular-like electronic structure by either shrinking the material to a size that is smaller than the de Broglie wavelength, where quantum and dielectric confinement impacts the structure,[109,110] or by reducing the connectivity of the octahedra.[111] Progressing from 3D perovskite phases to two-dimensional sheets, one-dimensional wires, and zero-dimensional isolated octahedra results in materials with more discrete energy levels and gradually shifts the optical transitions to higher energies (Fig. 5a). The low formation energy of the materials allows for a highly tunable and diverse range of optical properties (Fig. 5b and 5c). Section 6 reviews three primary structural changes that can dramatically modulate MHP optical properties: crystal transformation, precipitation, and intercalation.

## 5. Review of Optical Write/Electrical Read MHP Materials and Devices

### *5.1 Two-terminal optical write/electrical read materials and devices*

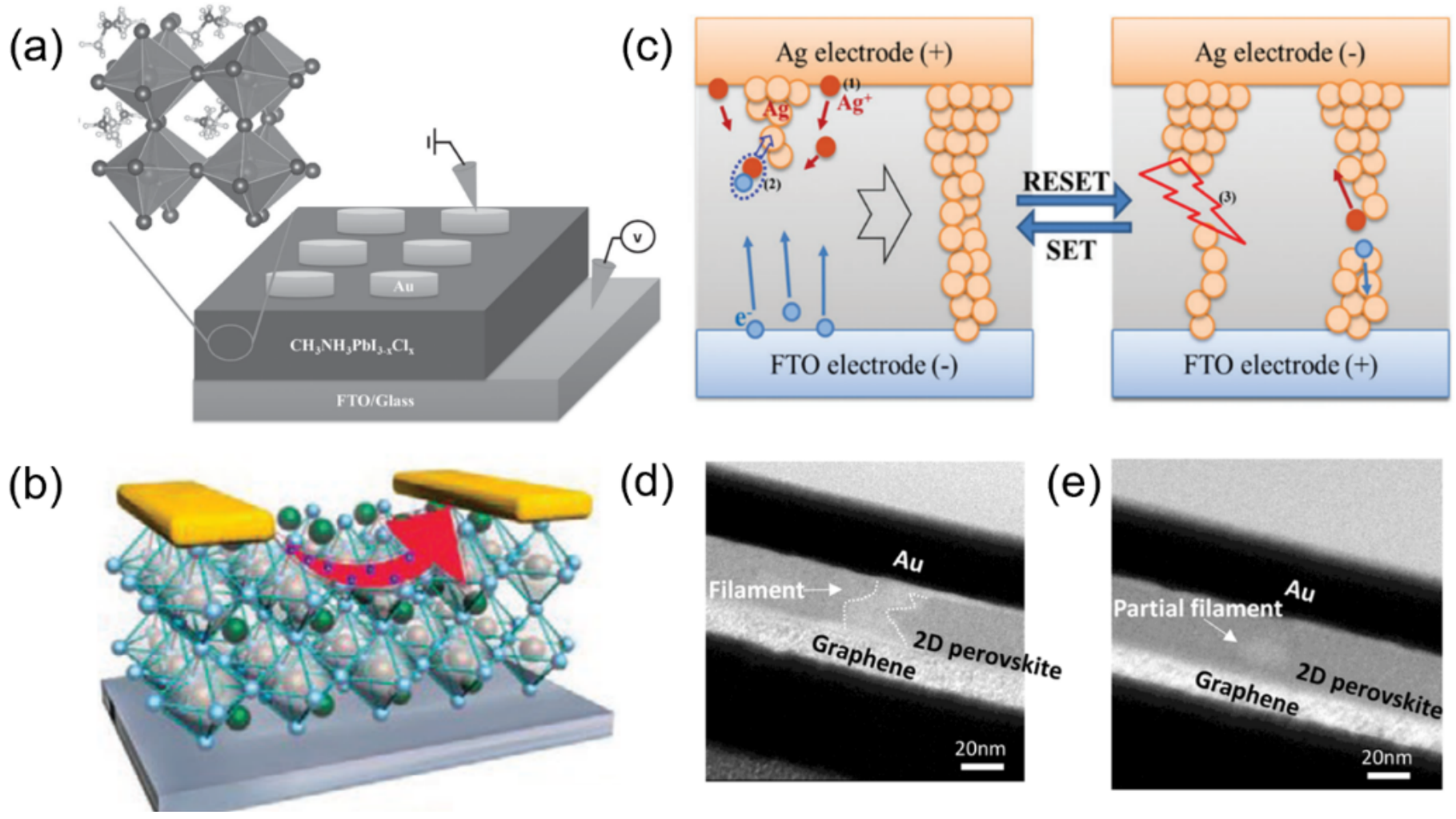

**Figure 6:** (a) Au/$CH_3NH_3PbI_{3-x}Cl_x$/FTO metal-semiconductor-metal structure reported by Yoo *et al.*[105] for RS. (b) Schematic of a lateral $MAPbBr_3$ synaptic device reported by Gong *et al.*[112] acquired with permission from reference[112]. (c) Mechanism for rupture and formation of a conductive filament by the participation of active metal (Ag) for setting and resetting RS, reproduced with permission from reference[106]. TEM image of (d) complete and (e) partial filament channel due to applied electric pulses, adapted from reference[21].

We begin by briefly summarizing early work on all-electrical MHP resistive switching (RS) devices that sets the stage for optically stimulated devices. Two-terminal devices can be arranged in either a layered metal-semiconductor-metal (MSM) structure (Fig. 6a) or in a lateral device geometry (Fig. 6b). An early all-electrical Au/$CH_3NH_3PbI_{3-x}Cl_x$/fluorine-doped tin oxide (FTO) MSM structure (Fig. 6a)[105] showed RS behavior consistent with a charge-trap controlled space-charge-limited conduction (SCLC) mechanism.[105] The devices showed endurance (write and erase cycles before failure) and retention (maintaining the information state even in the absence of power supply) for 100 cycles and $10^4$ s, respectively.[105]

This performance was further improved by replacing the gold electrode (presumed to be 'inert') with an 'active' silver electrode,[106] where a positive bias on the Ag electrode forms a conductive $Ag^0$ filament and negative bias ruptured the filament *via* Joule heating (Fig. 6c). Although this filamentary mechanism improves device on current, the electrochemical reactions can compromise MHP stability and reduce device lifetime.[113] These stability issues can be mitigated by an $MoO_3$ intermediate layer.[104] Instead of filament formation, the operational mechanism of Perovskite/$MoO_3$/Ag synaptic devices relied upon $Ag^+$ and $I^-$ migration to reversibly form/annihilate an AgI layer at the perovskite/$MoO_3$ interface.[114] Another study by Xu *et al.*[76] utilized a Al/$CH_3NH_3PbBr_3$/buffer-layer capped conducting polymer (BCCP) architecture with no active top electrode. Low-voltage short pulses controlled reversible ion migration and high-energy pulses caused ion accumulation at the MHP/BCCP interface to realize neuromorphic functionality. Similar results and mechanisms were reported on several other MHP based devices.[24,101–103]

Mixed ionic-electronic conductivities and large structural and electronic anisotropy suggest 2D MHPs as interesting candidates for synaptic emulation. Tian *et al.*[21] proposed a filamentary mechanism for realizing extremely low operating current (10 pA) resistive memory in a graphene/$(PEA)_2PbBr_3$/Au synaptic device that incorporated 2D MHP layers exfoliated from single crystals. Figures 6d and 6e show full and partial filaments (white dotted lines) formed by $Br^-$ vacancies. They summarized the mechanisms in three stages: (a) movement of the $Br^-$ ions due to a positive field applied at the Au electrode, (b) filament formation creating a conduction channel and (c) high conduction leading to Joule heating which eventually reduces the filament thickness. The energy consumption values of this device (400 fJ/spike) and more recent analogous 2D devices (e.g. $(PEA)_2(MA)_{n-1}Pb_nBr_{3n+1}$, 0.7 fJ/spike)[24] are quite low and approach the range of biological synapses (1-10 fJ/event).

While single-crystal exfoliation may provide high-quality MHP synapses, it may not ultimately be scalable. To address this, Gong *et al.*[112] developed a thickness-confined surfactant-assisted self-assembly approach to generate single-crystalline $CH_3NH_3PbBr_3$ platelets with controlled thickness and lateral dimensions. A lateral two-terminal device geometry (Fig. 6b) provided flexibility for the conduction channel length and hence the time scales for charge carrier transport and device outputs. The resulting devices demonstrated sub-pA range operating current, along with a vast array of synaptic behaviors.

Whereas all of the two-terminal devices discussed thus far were purely electronic, Ham *et al.*[115] studied the ability of photoexcitation to modulate the RS behavior of Ag/$CH_3NH_3PbI_3$/ITO/Glass devices (Fig. 7a). Light illumination accelerated the electric field-driven migration of iodide vacancies and filament formation resulting in below 0.1 V synaptic operation. The presence of an active metal Ag electrode suggests a working mechanism based on filament formation (Fig. 7b and 7c), where photogenerated charge carriers help in reducing the activation energy for ion migration and filament formation (Fig. 7d). The photogenerated electric field ($E_{ph}$) likely supported the applied electric field ($E_{Ext}$) in the device, lowering the activation energy ($E_a$) by a factor of $zeE/2$. Here, $z$ is the ionic charge, $e$ is the charge on an electron ($1.60 \times 10^{-23}$ C), and $E$ ($E_{ph} + E_{Ext}$) is the overall applied electric field. The $E_a$ might be lowered further if there are other mechanisms involved such as light-induced phase transition, charge accumulation at grain boundaries or lowered Schottky barrier.[115] The effect could be even more prominent if ferroelectricity and hence ferroelectric photovoltaic effect are present.[99] Further Ham *et al.*[115] investigated the learning capability of a 28 × 28 pixel array (784 input neurons) of their light-assisted MHP-based devices by simulating pattern recognition using the Modified National Institute of Standards and Technology (MNIST) database. They achieved 82.7% accuracy after 2000 learning phases *via* light-assisted revision of the synaptic weights based on long-term potentiation or depression of the synaptic devices. The power consumption for their light-assisted devices was 2600 times lower than the devices operating merely under electronic inputs.

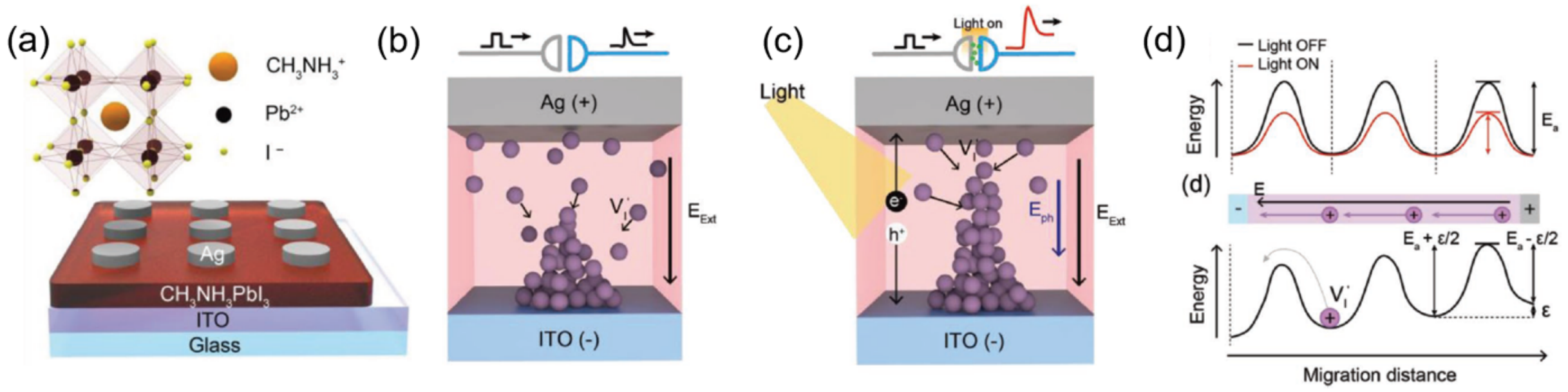

**Figure 7:** (a) Device design used by Ham *et al.*[115] for light-assisted synaptic operation and (b)-(c) its mechanism. The mechanism of filament formation in (b) dark and (c) under light illumination. $E_{Ext}$ is the applied electric field while $E_{ph}$ is the photogenerated field. (d) A variation in the activation energy ($E_a$) for filament formation (top) and migration distance of iodine vacancies. Reproduced with permission from[115].

*5.2 Three-terminal optical write/electrical read materials and devices*

Simultaneously performing signal transmission and learning processes is a prerequisite for neuromorphic computing in artificially engineered synapses. Meeting these requirements is often difficult to achieve with two-terminal devices, but can be facilitated by three-terminal devices consisting of gate, source and drain connections. The pioneering work on a traditional three-terminal FET based on MHP was reported by Kagan *et al.*[69] at IBM. They explored a 2D perovskite ($(C_6H_5C_2H_4NH_3)_2SnI_4$, also known as $(PEA)_2SnI_4$) and demonstrated a high on-off ratio ($10^4$). Much additional work has been performed on MHP FET devices[116–118] and their relevant charge transport mechanisms[119] in recent years. In three-terminal *optically addressed* devices, the traditional electrical gate *and* an optical/optoelectric stimuli can modulate the conduction channel between source/drain electrodes. An early example is the demonstration of a 3D MHP ($CH_3NH_3PbI_3$) phototransistor with a photo-response time of less than 10 μs and photosensitivity of 320 $AW^{-1}$.[120] This configuration sets the stage for optically stimulated *neuromorphic* devices that not only allow simultaneous signal transmission and learning but also provide the possibility of three-dimensional expansion for parallel data processing.[121–124] The anisotropic electronic conductivity in 2D MHPs, which favors in-plane transport over out-of-plane transport, would suggest that these MHPs would be well-suited to three-terminal devices. In such devices, MHP layer could either be used as an active channel or as a floating gate[125] which are discussed in the following sections in detail.

### *5.2.1. Three-terminal device with an MHP active channel*

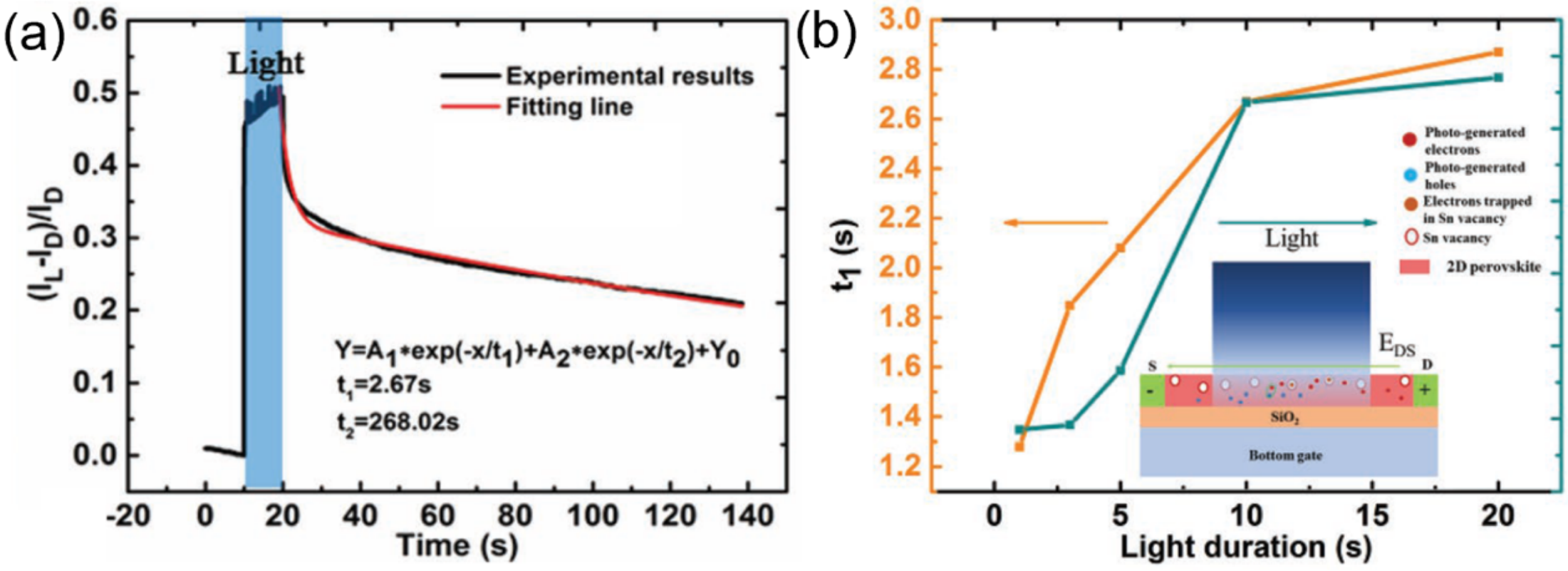

**Figure 8:** (a) Observation of photocurrent decay constants ($t_1$ and $t_2$) from experimental data in $(PEA)_2SnI_4$ based LSST, (b) The mechanism of charge trapping due to Sn vacancies and its relationship with time decay constants as a function of pulse duration. Time constants ($t_1$ and $t_2$) as a function of $SnF_2$ addition cause suppression of Sn vacancies and suggest $t_2$ is connected to the deeper traps while $t_1$ is related to shallow traps. Obtained with permission from reference.[126]

Zhu and Lu illustrated tunable synaptic functions in a two-terminal $Ag/CH_3NH_3PbI_3/Ag$ lateral device incorporating light illumination with variable wavelength and intensity as a gate.[127] The device was operated under pure electrical, purely optical, and a combination of opto-electric inputs. The device demonstrated synaptic coincidence detection, i.e., distinguishing optical and electric outputs occurring at the same time. Sun *et al.*[126] disclosed a light-assisted synaptic device based solely on a $(PEA)_2SnI_4$ conducting/active channel. Simple exponential fitting of the photocurrent decay (Fig. 8a) suggests the presence of two decay constants $t_1$ and $t_2$ which also depend on the light exposure duration (Fig. 8b). The authors hypothesized a possible mechanism of charge trapping due to Sn vacancies (inset Fig. 8b), with the two decay constants corresponding to two types of trapped states that were either near-surface (shallow) or bulk (deep). This hypothesis was tested by exposing the device to $SnF_2$, which suppressed near-surface Sn vacancies. A reduction in $t_2$ with $SnF_2$ exposure suggested that it is connected to the deeper traps while $t_1$ is related to shallow traps. This device showed PPF, STP, LTP and a transition between STP to LTP that could be controlled by incident spike frequency.[118]

*5.2.2. Three-terminal device with an MHP floating gate*

Another effective strategy for three-terminal optical synapses involves the utilization of heterostructures that can potentially decouple the various steps required to realize synaptic behavior - e.g., large absorption, charge trapping and release, good lateral mobility and conductance, etc. An early demonstration of this type of heterojunction phototransistor approach was shown by Wang *et al.*[61] for $CsPbBr_3$ MHP quantum dots (QDs). Figure 9a shows their photonic flash memory or light-stimulated synaptic transistor (LSST) device structure: (Si/$SiO_2$/$CsPbBr_3$ QDs/polymethyl methacrylate (PMMA)/pentacene/Au). In this configuration, the strong optical absorption in the $CsPbBr_3$ combines with the good lateral conductivity in the pentacene semiconducting layer to facilitate optical programming and electrical erasing of the memory device. A type II band alignment between $CsPbBr_3$ and pentacene layers dissociates photogenerated excitons at the interface, while a very thin PMMA layer between the QD and pentacene layers hinders charge recombination. The trapped charges were suggested to be released as a consequence of the electrical inputs *via* electrodes. Figure 9b explicates the mechanism, with the help of a schematic energy band diagram, for the programming and erasing operations. Figure 9c-e demonstrates photonic potentiation using variable input light wavelength, intensity and pulse durations. Notably, the device was able to return to its original state with a pulse of -50 V applied for 1 s. With a combination of optical inputs and electrical erasing they were able to demonstrate STP, LTP, paired-pulse facilitation (PPF), paired-pulse depression (PPD), and SRDP. They claimed to achieve low energy operations of 1.4 nJ/event. Interestingly, illumination controlled/assisted operations depend on the absorbed energy from the incident wavelength suggesting that materials capable of absorbing longer wavelengths will be relatively more energy efficient.

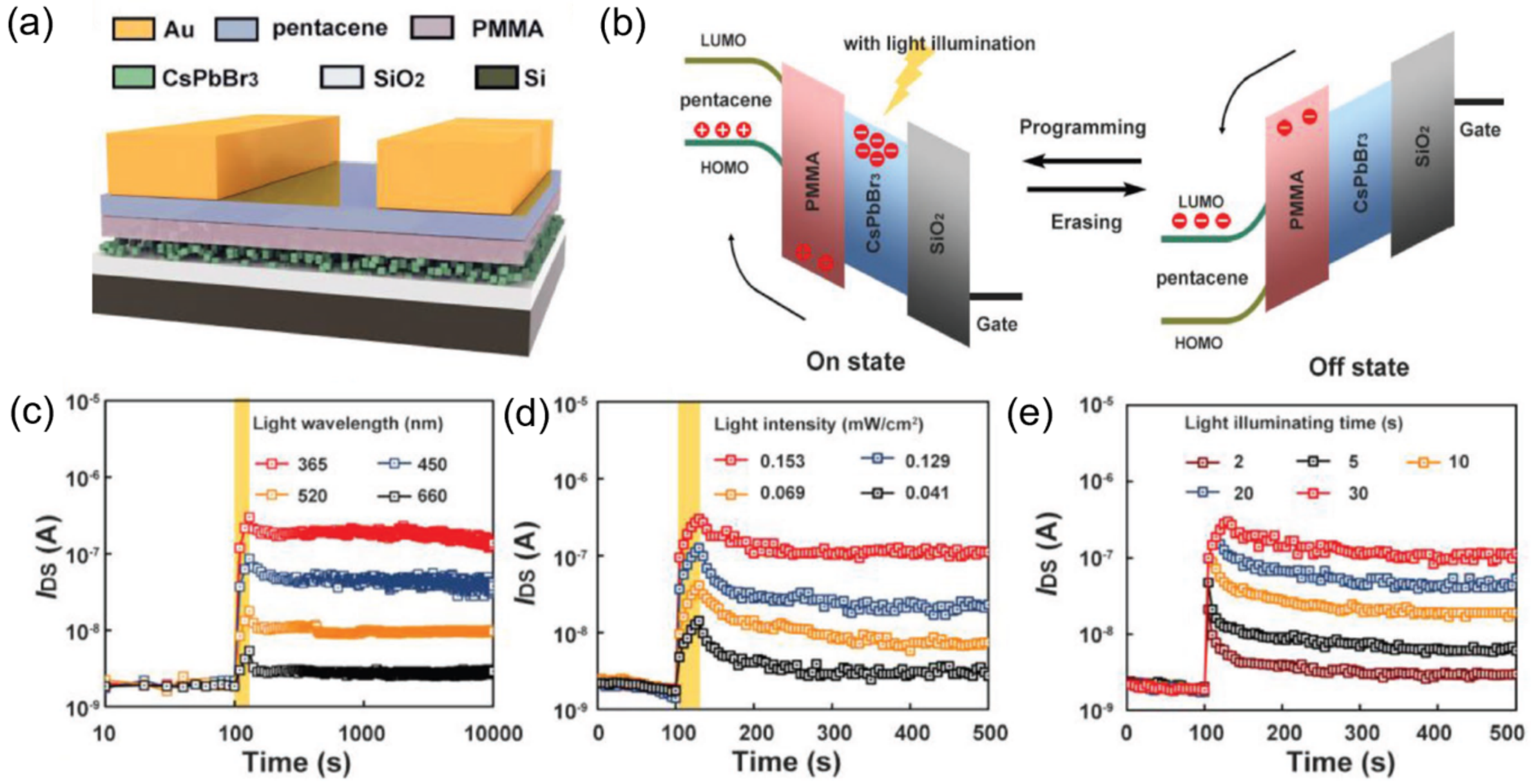

**Figure 9:** (a) Schematic of photonic flash memory reported by Wang *et al.*[61]. (b) mechanism of optical trapping and electrical releasing of the charges. Source-Drain current ($I_{DS}$) as a function of (c) input wavelength for light pulse (30 s) with a fixed intensity of 0.153 mW cm$^{-2}$, (d) light intensity for pulses of 30 s each at a fixed wavelength of 365 nm, (e) pulse duration corresponding to a fixed wavelength of 365 nm with an intensity of 0.153 mW cm$^{-2}$. The yellow shadow indicates the duration of the light pulse. Reproduced with permission from reference[61].

Wang and co-workers[64] used a similar heterojunction approach by blending $CsPbBr_3$ QDs with the organic semiconductor (poly(3,3-didodecylquarterthiophene), PQT-12) to achieve photoinduced charge separation and delayed photocurrent decay. The delay is suggested to be only possible due to the PQT-12 blend and was absent in pure $CsPbBr_3$ QDs. The phenomenon of delayed decay is identical to a forgetting process of the human brain. Their LSST devices also exemplified EPSC, high-pass dynamic filter, PPF, memory, and learning behaviours. Hao *et al.*[128] employed the organic semiconductor poly[2,5-(2-octyldodecyl)-3,6-diketopyrrolopyrrole-alt-5,5-(2,5-di(thien-2-yl)thieno [3,2-b]thiophene)] (DPPDTT) with $CsPbBr_3$ quantum dots to simulate synaptic behaviours such as EPSC, PPF, a transition from STP to LTP, and learning behaviour experience. Their devices showed a variation in drain current (2-15 nA) with increase in number of optical pulses. No gate voltage (only light illumination with intensity of 0.26 $mWcm^{-2}$) was applied and the source-drain voltages were maintained at -0.2 V. In addition, operation with a weak response at low energy (at $V_G$ = 0 V, $V_{S\text{-}D}$ = -0.0005 V, λ = 450 nm (0.05 $mWcm^{-2}$), 0.5 fJ/spike) was also demonstrated. Moreover, the magnitude of the EPSC response showed negligible decay on re-testing after storing the devices in the glovebox for over 1 year. They also achieved 'AND' and 'OR' logic functions by regulating the input synaptic parameters.

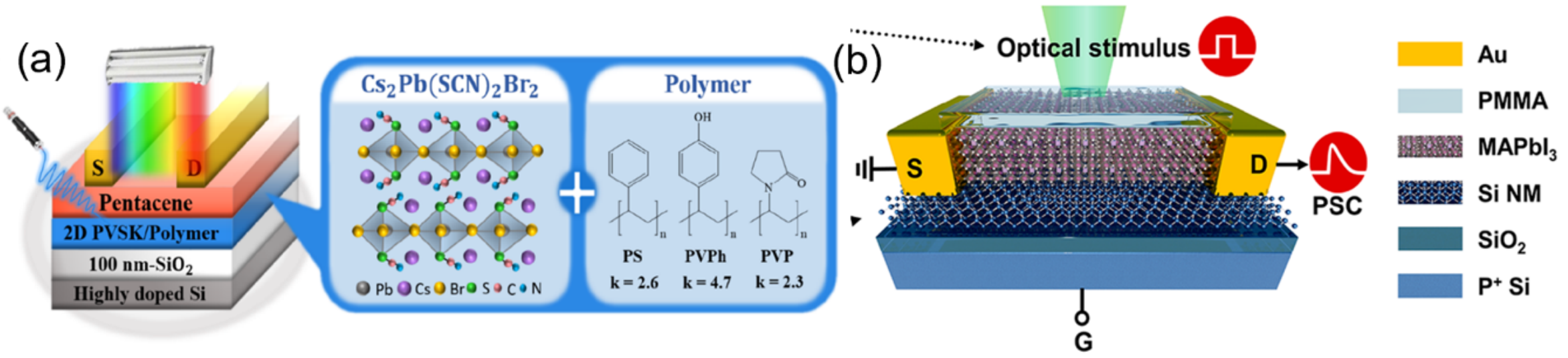

**Figure 10:** (a) Device configuration of photo-memories based on $Cs_2Pb(SCN)_2Br_2$ and the host polymers (polystyrene, poly(4-vinylphenol), and poly(vinylpyrrolidone)). Reproduced from reference[129]. (b) Schematic of the synaptic transistor design based on Si nanomembrane and $MAPbI_3$. Adapted from reference[130].

Recently, Liao *et al*[129] manifested an optically stimulated heterojunction memory device using a pentacene conducting channel in contact with a blend of 2D MHP $Cs_2Pb(SCN)_2Br_2$ nanocrystals within an insulating (large-bandgap) polymer host (Fig. 10a). Three polymer hosts were compared for the devices – polystyrene (PS), poly(4-vinylphenol) (PVPh), and poly(vinylpyrrolidone) (PVP). The devices were fully optically controlled, and the envisaged mechanism was based on charge trapping and de-trapping within the MHP/polymer layer. In particular, white light illumination (primarily exciting the MHP) can be used to write hole photocurrent in the pentacene channel, whereas a 450 nm blue laser (primarily exciting pentacene) erased the photocurrent. While the study did not demonstrate specific synaptic functions, it motivates investigating devices based on MHP hosted in polymer layers for all-optical neuromorphic operation, especially the role of the inert polymer host.

Recently, a back-gated device design was reported by Yin *et al.*[130] The gate electrode was covered by a silicon nanomembrane followed by a $MAPbI_3$ layer and top source/drain contacts (Fig. 10b). Source-drain current flowed preferentially through the high-mobility Si nanomembrane, while photon absorption occurred predominantly in the $MAPbI_3$ layer, due to its much larger absorption coefficient. Photogenerated charge carriers in $MAPbI_3$ control the built-in potential in the silicon nanomembrane which in turn governed the post-synaptic outcome at the drain. The gate voltage provided an additional degree of freedom over synaptic weights and allowed the authors to mimic visual learning and memory under distinct mood states. The device was operated under a low drain voltage of 0.1 V and an optical power density of 1 $\mu W/cm^2$, and capable of emulating EPSC, PPF and STP to LTP transition.

Hao *et al.* prepared photo-transistor devices based on heterojunctions between perovskite nanocrystals ($CsPbBr_3$, $FAPbBr_3$, and $CsPbI_3$) and highly enriched semiconducting single-walled carbon nanotubes (s-SWCNTs) that were dominated by the (6,5) chirality (Fig. 11a).[59] In these devices, the channel current was carried by the high-mobility s-SWCNTs, while the MHP nanocrystals (NCs) served as the dominant light absorber. Photocurrent was driven primarily by photoinduced hole transfer, from NCs to s-SWCNTs, across the Type-II interface. Photoexcitation of the devices at 405 nm or 532 nm, predominantly exciting the MHP NCs, led to a combination of a prompt and delayed source-drain photocurrent at room temperature. For both continuous and pulsed excitation, the photocurrent decayed slowly, over the course of hundreds to thousands of seconds. A combination of spectroscopic, temperature-dependent photocurrent, and time-of-flight secondary ion mass spectrometry (TOF-SIMS) measurements was used to analyze the mechanisms contributing to the observed prompt and persistent photoconductivity. The authors concluded that a combination of charge trapping and ion migration in the nanocrystal layer, aided by the photoinduced field (orthogonal to the transport channel), contributed to the long-lived photocurrent in the s-SWCNT transport channel. Low-fluence measurements demonstrated exceptionally low switching energies for the optical memory at no applied gate voltage ($V_G$ = 0 V), in the range of 84 pJ/pulse (total pulse energy delivered to device) or 75 fJ/pulse (pulse energy incident on device area). The optically stimulated synaptic weight could be adjusted systematically by applying small 300 ms 10 V gate voltage pulses (Fig. 11b) or could be completely erased by a larger 20 V gate pulse. Spike-frequency dependent plasticity was demonstrated for 30 μs 405 nm pulses over a frequency range of 0.1 – 100 Hz (Fig. 11c).

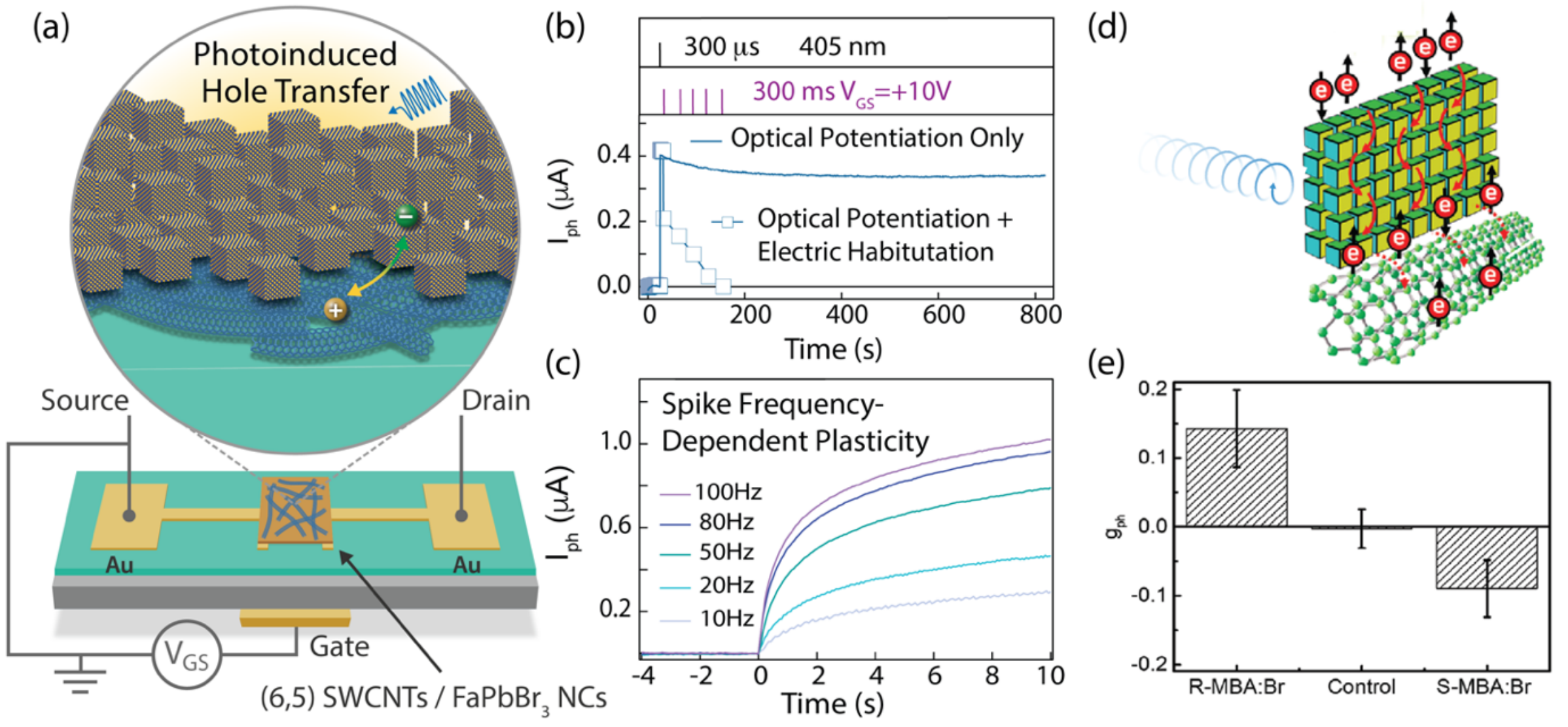

**Figure 11. (a)** Schematic of three-terminal optical synapse that generates persistent photocurrent via photoinduced hole transfer from MHP NCs to semiconducting SWCNT network transport channel. **(b)** Demonstration of optical potentiation of device in (a) with a 300 μs 405 nm pulse. The photocurrent 'weighting' can be adjusted by a series of 300 ms gate voltage pulses ($V_{GS}$ = +10V). (c) Demonstration of spike-frequency dependent plasticity in the device in (a). Adapted with permission from ref. *[59]* **(d)** Similar three-terminal device where the $CsPbBr_3$ NCs are modified with chiral ligands to produce synaptic weighting that depends on the circular polarization of incident photons. **(e)** Photocurrent anisotropy (g-factor, $g_{ph}$, Eqn. 1) for devices prepared with chiral R- or S-MBA ligands or non-chiral octylammonium ligands. Adapted with permission from ref. *[60]*

The same authors subsequently demonstrated that the polarization of incident photons can provide another degree of freedom in these heterojunction optical synapses.[60] The same NC/s-SWCNT three-terminal devices were studied, but in this case, the standard octylammonium ligands on the $CsPbBr_3$ MHP NCs were replaced by chiral ligands (R- and S-methylbenzylammonium, MBA). Excitation of the R-MBA capped NCs with right-handed circularly polarized (RCP) pulses generated larger photocurrent ($I_{ph,R}$) than that generated by left-handed circularly polarized (LCP) pulses ($I_{ph,R}$), while this anisotropy was reversed for s-MBA capped NCs. The photocurrent anisotropy of such devices is typically quantified by the so-called g-factor:

$$g_{ph} = 2 \times \frac{I_{ph,R} - I_{ph,L}}{I_{ph,R} + I_{ph,L}} \qquad \text{Equation 1}$$

The g-factor of these MHP NC/s-SWCNT optical synapses was high, in the range of +0.14 for R-MBA devices and −0.1 for s-MBA devices, in line with the anisotropies found for a similar 0D copper iodide-based heterojunction.[131]

In a similar attempt, Liu *et al.*[132] explored double perovskite $CsBi_3I_{10}$ transistors and reported a high responsivity (($6.0 \times 10^4$ A $W^{-1}$)) and detectivity ($2.5 \times 10^{14}$ jones). However, they could only achieve an on-off ratio of $10^2$. A similar order of on-off ratio was reported for $Cs_2NaBiI_6$ [133] which was substantially improved to $10^7$ using heterostructures between the monocrystalline double perovskite $Cs_2AgBiBr_6$ and Indium-Gallium-Zinc oxide (IGZO).[134] Superior absorption of the visible light spectrum and effective charge transfer across the $Cs_2AgBiBr_6$ and IGZO interface [134] helped achieve a better photo-response and plasticity than the devices comprised of only an IGZO layer.[134] The $Cs_2AgBiBr_6$/IGZO devices illustrated several synaptic behaviors including EPSC, PPF, STP, and LTP along with a pattern recognition accuracy rate of 83.8±1.4%. Cao *et al.*[135] used $CsPbI_2Br$ perovskite nanocrystals with IGZO in the same floating gate arrangement and reported a high photoresponsivity ($4.18 \times 10^5$ $AW^{-1}$) and detectivity ($1.97 \times 10^{17}$ Jones). Their device was documented to consume 2.6 pJ/light spike for detectable ΔEPSC. A study by Periyal *et al*[136] utilised $CsPbBr_3$ quantum dots with IGZO semiconductor active layer to realize a two-terminal device with light illumination acting as a gate for controlling the synaptic weights. They also illustrated electro-optic gating-controlled conductance states.

*5.3 Higher-order arrays with neuromorphic functionality*

Once the characteristics of a device are well understood and modelled, arrays of devices could be integrated to perform computing tasks. In this context, John *et al.*[101] provided an early demonstration of all-electrical drift-diffusive synapses for neuromorphic computation using two-terminal memristive synaptic devices based on $CH_3NH_3PbBr_3$, $CH(NH_2)_2PbBr_3$, and $CsPbBr_3$ MHP layers. The device response was modulated by the impact of carrier injection and migration of anion and cation vacancies on the carrier injection barriers at the interfaces with the electron and hole transport layers (ETL and HTL). Understanding of this ionotropic behaviour was used to fabricate an array of 4 x 4 devices to represent 16 pixels, which could be trained to recognize, retain, and/or forget letters such as 'N', 'U', and 'T'. A two-layer neural network model was proposed for the recognition of hand-written numbers.

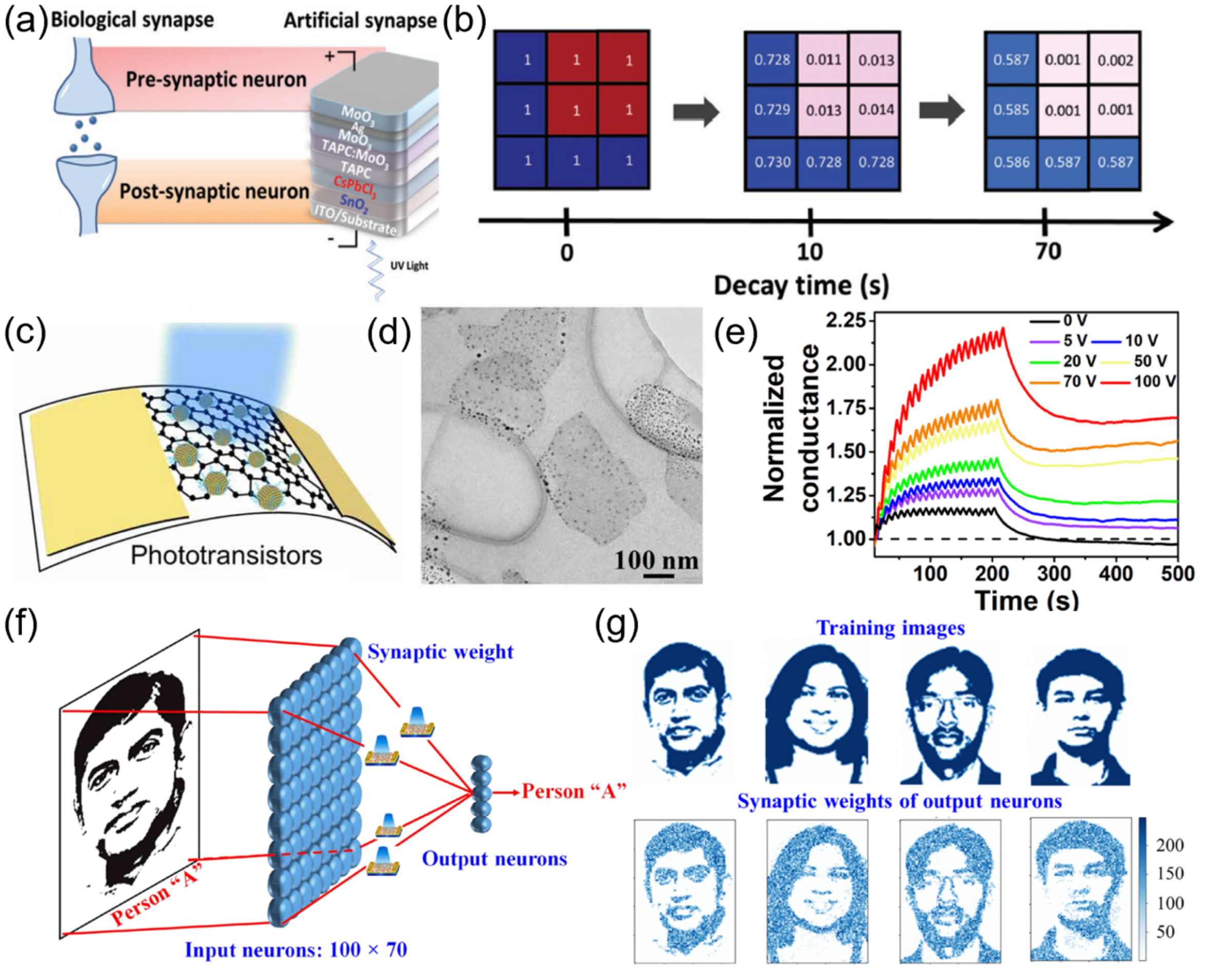

**Figure 12:** (a) Schematic of the pre- and post-synaptic neuron device layers. Adapted from reference [137]. (b) Decay measurement (note the difference in color contrast) of pixels after 10 s and 70 s. The numbers shown in the boxes are normalised post-synaptic current values. Reproduced with permission from reference [137]. (c) Schematic of MHP QDs on graphene superstructure photo-transistor (d) TEM image of MHP QDs grown on a single layer of a graphene sheet. (e) Gate-dependent transient characteristics of the device after application of 20 optical pulses. (f) 100 x 70 neuron network structure for face recognition. (g) Images used for training (top) and recognised images and the synaptic weights corresponding to the output neurons (bottom). Reproduced from reference [58].

Yang and co-workers[137] introduced transparent and flexible inorganic perovskite two-terminal photonic synapses, with a response based on light-illumination induced charge trapping and de-trapping in the $SnO_2$-$CsPbCl_3$ interface (Fig. 12a). The devices were integrated into 3 x 3 arrays for recognition of the letter ‘U’ and 'L'. Figure 12b shows the decay of 'L' shaped pixels (written in blue and red) after 10 s and 70 s. The numbers shown in the boxes are normalised post-synaptic current values. Along similar lines, Ma *et al.*[138] explored perovskite nanoplates of inorganic $CsPbBr_3$ in a two-terminal ITO/PEDOT:PSS/CuSCN/$CsPbBr_3$/Au laminated structure. Trapping and de-trapping of charge at the PEDOT:PSS/CuSCN interface was credited for conductance changes, whereas the optical response was attributed to the $CsPbBr_3$ perovskite nanoplates. The devices were arranged in arrays of 3 x 2 patterns and were written with UV light sources of different energies. The arrays were used to demonstrate learning, forgetting and backtracing of the optically written patterns with low voltages of 0.1V, 0.25 V and 0.5 V respectively.

In a recent study, Pradhan and colleagues[58] explored photoactive $CH_3NH_3PbBr$ QDs, grown on monolayer graphene conducting channel, in a phototransistor configuration (Fig. 12c-d). 20 illumination pulses at 430 nm tuned the conductance state of the device, which was maintained after the removal of illumination if a gate voltage was applied (Figure 12e). The photonic synapse on these three-terminal devices was observed to operate as low as 36.56 pJ per spike (photon energy incident on device area). These characteristics were used to design a neuron network of 100 x 70 for facial recognition (Fig. 12f-g), a strategy suggested as translatable to other 2D materials for energy-efficient and relatively fast computation. In a more recent attempt, Lee *et al.*[139] explored nanocones of $CH_3NH_3Br_3$ MHPs templated in self-assembled block copolymers. They took advantage of the off-center spin-coating process to achieve spatially distributed MHP cones to emulate the functioning of a human retina. Authors formed an array of 60 x 12 to show device position dependent functioning as receptor and synapse. Their devices illustrated MNIST pattern recognition efficiency of 90 %.

## 6. Optical Read: Phase transitions in MHPs for optical switching and memory

Optical read in memory applications avoids electro-optical conversion and undesired Joule heating, which avoids the speed and heat dissipation problems associated with electrical memory. Free-space, on-chip, and fiber-integrated optical memories often rely on phase-change materials that transition from a high absorbance/reflectance state to a low absorbance/reflectance state due to photothermal heating of the material. For the prototypical phase-change chalcogenide ($Ge_2Sb_2Te_5$), it is a crystalline-to-amorphous transition that enables state tuning.[140] Neuromorphic synapse devices require more than two distinct optical states. Conventional PCMs have shown multilevel operation by varying optical pulse energy[141] and have demonstrated multilevel synaptic behavior when multiple PCMs are positioned in series in an on-chip photonic synapse.[142] Both examples illustrate how materials with two discrete optical states can be leveraged for synaptic behavior. Neuronal “integrate and fire” behavior is also observed in two-state materials where partial state transformations are achieved in each excitation event (e.g. optical pulse) and the succession of such events leads to crystallization and complete state transformation.[143]

The following section describes switching between two or more optical states in MHP materials. Though we highlight some examples of memory applications, the purpose of this section is to demonstrate the unexplored potential of MHP materials as optical memory and neuromorphic materials based on switchable optical properties. We review three distinct mechanisms of control over MHP octahedra connectivity and resulting optical properties: crystal transformation, precipitation, and intercalation. Each mechanism introduces control in optical properties that can span the ultraviolet to near-infrared bands of the electromagnetic spectrum, depending on the specific chemistry (Fig. 5). We focus this section on mechanisms that fit the following criteria: (1) large changes in optical density or emission, (2) reversibility, and (3) rely on one or fewer chemical stimulus. For instance, though ion exchange is a common method to reversibly tune the optical properties of perovskite materials,[144] especially X-site anion exchange in nanomaterials,[145] it typically requires two separate chemical stimuli to achieve the forward and reverse ion exchange reaction. The same is true for transformations of the Cs-Pb-Br NC system,[146] where two chemical stimuli will reversibly transform 3D $CsPbX_3$ materials into 0D $Cs_4PbX_6$.[147][148]

We are sensitive to the notion that the definition of "perovskite" is often abstracted from its original definition of a 3-dimensional corner-sharing metal halide octahedral networks.[149] Each transformation considered here has at least one state that has a 2-dimensional perovskitoid with corner-sharing octahedra or 3-dimensional perovskite but may transition to other non-perovskite structures. We focus this section on Pb- and Sn- based materials due to their high prevalence and diversity of switching mechanisms found in the literature. A discussion on switchability in materials not based on Pb or Sn is included in the future outlooks and perspectives section (Section 7).

### *6.1 Crystal transformations*

Two categories of crystal transformations lead to large swings in optical density: crystal phase transition and crystalline-amorphous transformations. Lower dimensional Pb- or Sn-based halide perovskite or perovskitoid materials exhibit changes in optical properties due to reversible crystal phase transitions induced by changes in pressure or temperature. For example, 1D $PbI_4 \cdot$4-MAPY (4-MAPY = pyridin-4-ylmethanamine) exhibits reversible red-shifted visible absorption when heated above 420 K.[150] A similar absorption shift occurs in 2D $(MA)_2[PbI_2(SCN)_2]$ when under applied pressure.[151] 2D [methylhydrazinium]$_2PbI_4$ has a bandgap of 2.2 eV at room temperature, but absorption and emission blueshift with decreasing temperature.[152] In each example, the connectivity of the octahedra remain intact. Changes in optical properties occur due to octahedral tilting and bond distortions, which alters the coupling between octahedra but does not result in change in octahedra connectivity.

Octahedral tilting and distortion in 3D perovskite materials are more constrained and do not yield changes in optical properties as large as the lower dimensional analogs. The most widely studied 3D MHP is $MAPbI_3$. At temperatures >327 K, the structure of $MAPbI_3$ is represented by the cubic space group $Pm\bar{3}m$, and below 327 K, it exists in the tetragonal space group $I4/mcm$. The cubic-tetragonal transition is second order and continuous over 165–327 K.[153] The transition results from octahedral tilting and does not yield significant changes in optical properties, as both structures maintain 3D corner-sharing connectivity within the lead iodide lattice.

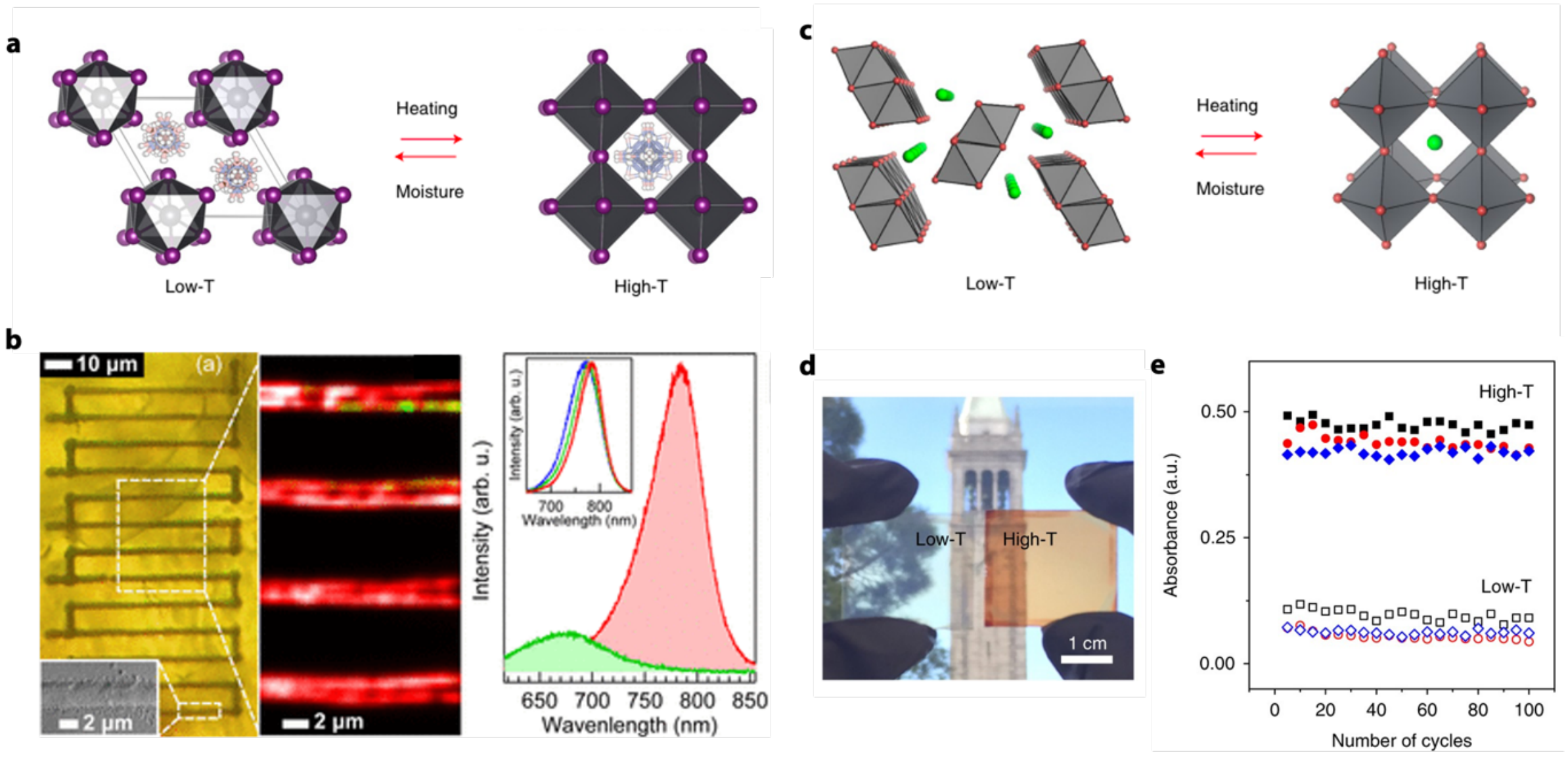

**Figure 13:** (a) δ- to α-Phase Transformation in formamidinium lead iodide to produce a shift in the absorption and emission properties of 0.4 eV. Adapted from ref. [154]. (b) Reversible direct laser writing of δ-phase FAPI to α-Phase FAPI [155]. (c) δ- to α-Phase Transformation in cesium lead iodide due to heat and moisture. (d) images showing the δ-phase (bleached) and α-Phase (colored). (e) Absorbance as a function of number of bleached-to-colored cycles of $CsPbI_2Br$ film. Adapted from Ref. [84].

Larger changes in optical properties occur when octahedral connectivity is altered. Replacing methylammonium with formamidinium produces thermodynamically favored non-perovskite δ-$FAPbI_3$ hexagonal phases with 1D face-sharing octahedra and 1.9 eV bandgap.[156][154] The 1.5 eV bandgap α-phase is stable at higher temperature (~170 °C) and metastable at room temperature (Fig. 13a), reverting to δ-phase after exposure to humidity. This metastability can be leveraged for producing two non-volatile, optically distinct states. Initial work showed reversible α- to δ-phase transformations by placing on a hotplate at 170 °C and then exposing to humid air.[157] Recent work showed the transformation can be achieved using direct laser writing with 458 nm continuous-wave-laser light to produce α-phase regions within a δ-$FAPbI_3$ matrix (Fig. 13b).[155] The different crystalline domains have features <2 μm, and the α-phase exhibits NIR emission that remains stable until exposure to moisture (Fig. 13b). Strain alters the phase diagram to favor the α-phase at room temperature,[156] and transition temperature is significantly decreased by growing strained nanocrystals in an oxide matrix, an effect realized by simply changing the vapor environment around the crystal at room temperature.[108]

Similar to the formamidinium system, replacing methylammonium with cesium produces a thermodynamically favored non-perovskite δ-$CsPbI_3$ orthorhombic phase.[158] For more details, the reader is directed to an extensive review on crystal phase transformations in $CsPbX_3$ materials.[145] Similar to $FAPbI_3$ analogs, the high-T phase is metastable at room temperature. Exposure to moisture or alcohols will also return $CsPbX_3$ materials to the orthorhombic δ-phase (Fig. 13c). The effect has been leveraged to produce thermochromic photovoltaic windows (Fig. 13d), which exhibit switchability over >100 bleached-to-colored cycles (Fig. 13e).[84] Pure FA and Cs-based lead iodide materials exist on the extreme ends of the tolerance factor, a phenomenological relationship between ionic radii that correlates to thermodynamic stability of halide perovskites in the cubic or pseudo-cubic phase.[159] A common strategy to stabilize the pseudo-cubic phase of perovskite materials is to alloy A-site cations, such as FA or Cs, to achieve a thermodynamically stable phase.[159] Unlike $FAPbI_3$, smaller X-site anions make $CsPbX_3$ materials more stable in the cubic or pseudo-cubic phases. X-site alloying with Br will effectively tune the δ- to α-phase transition. For $CsPbI_{3-x}Br_x$ ($0 \leq x \leq 3$), the phase transition temperature decreases nearly linearly from 320 to 105 °C as $x$ increases ($0 \leq x \leq 2.5$).[84] Quantum size effects have also been used to stabilize $CsPbI_3$ at room temperature.[109,160]

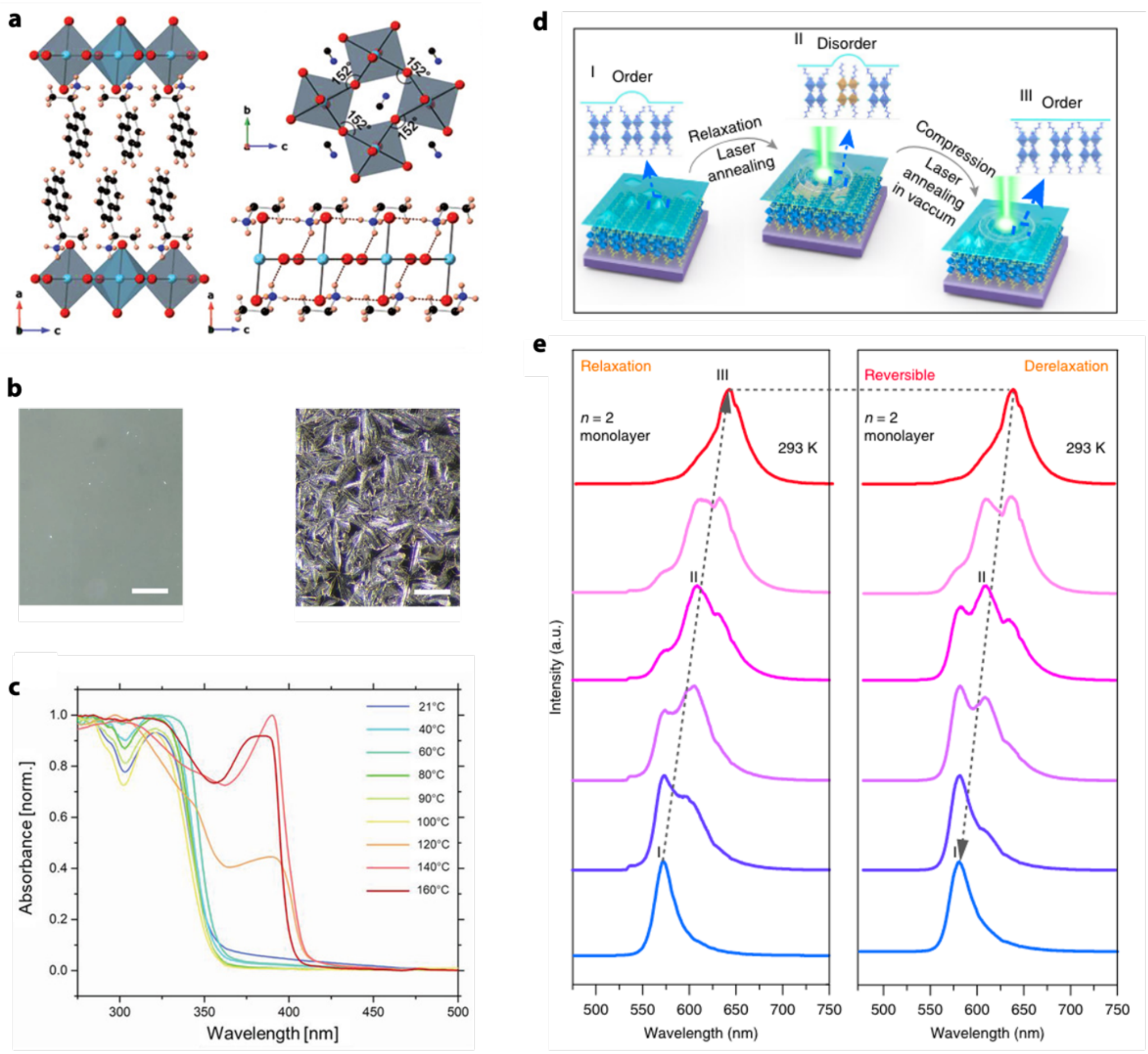

**Figure 14:** (a) Schematic of [*S*-(−)-1-(1-naphthyl)ethylammonium]$_2$$PbBr_4$ (*S*-NPB) crystal. (b) Optical images of amorphous (left) and crystalline (right). Scale bare 250 µm. (c) Corresponding normalized absorbance spectra at increasing temperatures. (d) Schematic of monolayer perovskite order-disorder transitions in $(CH_3(CH_2)_3NH_3)_2(CH_3NH_3)_{n-1}Pb_nI_{3n+1}$ where n=2. (e) Photoluminescience spectra of monolayer $(CH_3(CH_2)_3NH_3)_2(CH_3NH_3)_{n-1}Pb_nI_{3n+1}$ where n=2 [161].

Crystal-to-amorphous phase transformations have been leveraged for optical memory for decades,[140] as exemplified by the benchmark Ge–Sb–Te family of materials, which exhibit a metastable amorphous state that is obtained by rapid melt quenching.[162] 2D perovskitoid materials composed of chiral spacer molecules, [*S*-(−)-1-(1-

naphthyl)ethylammonium]$_2$PbBr$_4$ (*S*-NPB), demonstrate an analogous phase transformation with a lower melting temperature of $T_m$ = 175 °C than conventional materials (Fig. 14a,b). The transformation induces a shift in absorption onset of nearly 50 nm (Fig. 14c).[163] The material was designed to have a melting temperature below the decomposition or volitization temperature of the organic spacer ligand. Recent work synthesized pure-phase single-crystal $(CH_3(CH_2)_3NH_3)_2(CH_3NH_3)_{n-1}Pb_nI_{3n+1}$ ($n$ = 1–4) perovskitoid materials and exfoliated single quantum well layers.[161] Laser annealing induced a disordered state that could be reversed to a crystalline state by laser annealing in vacuum (Fig. 14d). The amorphous state exhibited a photoluminescence peak that is bathochromically shifted by ca. 100 nm from the crystalline state emission. In this case, exfoliated layers were encapsulated by 2D hexagonal boron nitride monolayers to prevent degradation or volatilization of the butylammonium spacer molecule, which demonstrates MHP-based heterostructuring as a powerful method to develop new functionality.

## 6.2 *Precipitation*

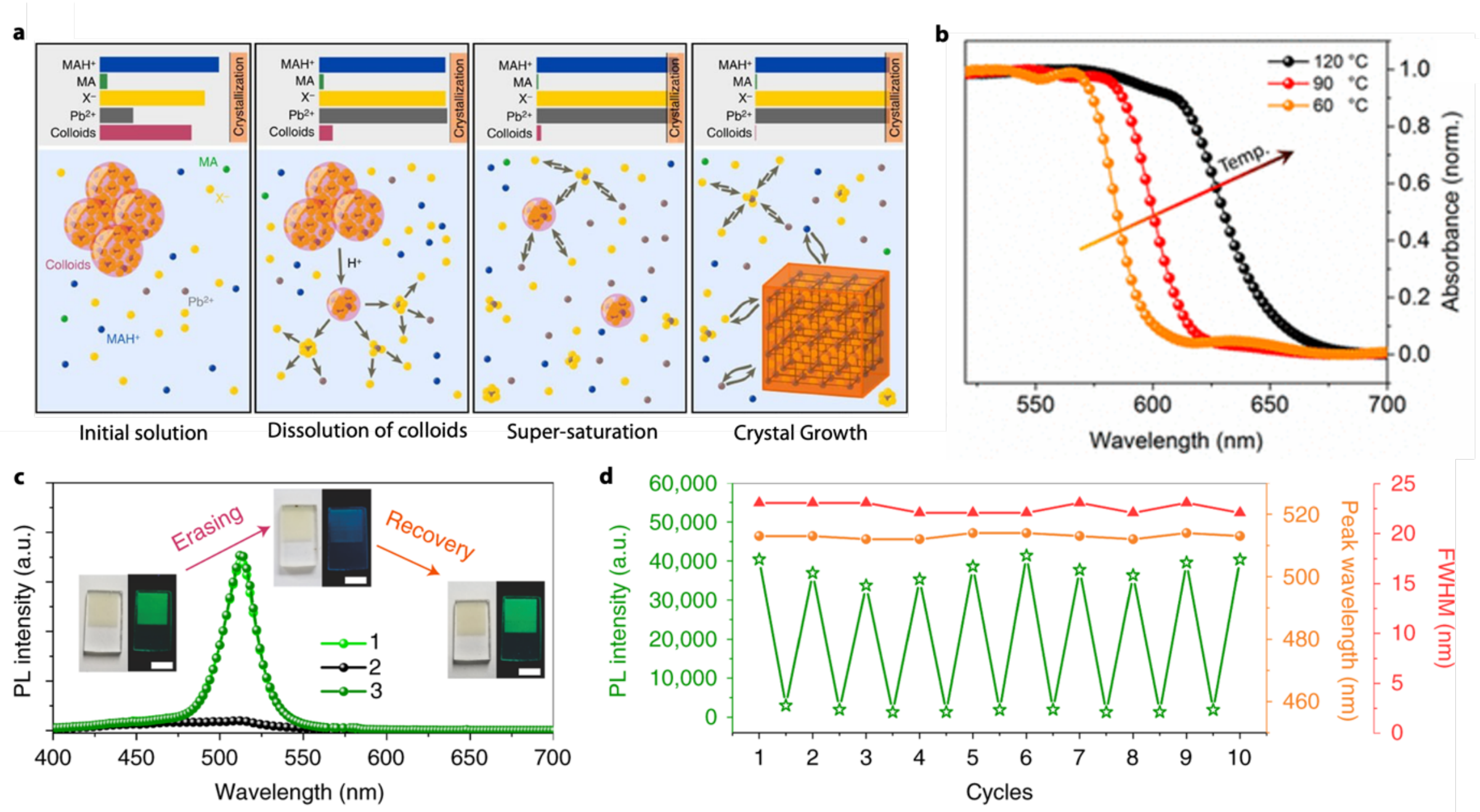


**Figure 15:** (a) Illustration of the mechanism of inverse temperature crystallization due to acid-base equilibria in the precursor solution [164]. (b) Application of inverse temperature crystallization to produce thermochromic inks [82]. (c) Laser writing and erasing of $CsPbBr_3$ nanocrystals in a glass matrix. (d) PL intensity, peak wavelength, and FWHM as a function of laser writing/erasing cycle [165].

We define precipitation reactions as those where an MHP material is reversibly formed and dissolved from an unstructured precursor matrix. Precipitation occurs upon supersaturation of precursors in the matrix. For MHP materials, supersaturation has been suggested to result from acid-base equilibria, where increasing temperature can change an acid–base equilibrium that raises the concentration of solute species due to dissolution of colloidal species to produce MHP crystals (Fig. 15a).[164] The effect has been named inverse temperature crystallization[166] and has been leveraged to produce thermochromic inks with temperature-dependent absorption onset (Fig. 15b). When the temperature is increased, nanocrystals are precipitated in solution and grown with increasing temperature,[82] and the nanocrystals are redissolved into the surrounding solvent upon lowering the temperature.

Reversible precipitation has also been demonstrated for $CsPbBr_3$ precursors in a glassy borosilicate-based solid matrix.[165][167][168] Femtosecond laser irradiation precipitates $CsPbBr_3$ nanocrystals with strong and stable photoluminescence (Fig. 15c) that are redissolved into the matrix by successive low-temperature annealing. The method was used to demonstrate ten write/erase cycles and writing schemes that yielded 2D lines, dots and patterns, and even 3D strucutres within the glass matrix using a three-axis translation stage.[167]

### *6.3 Intercalation*

The reversible mechanism offers the largest dynamic range in optical properties over the most chemically diverse range of MHPs. Here we arrange the discussion by the intercalant species and discuss intercalants in order of their increasing chemical complexity: (1) neutral molecules, (2) ions, (3) salt pairs, and (4) solvated salt pairs.

#### *6.3.1 Neutral molecule intercalation*

Neutral molecule intercalants are stabilized by molecular interactions that are significantly weaker than the ionic bonds associated with electrochemical ion intercalation. In 2D perovskitoid materials, the functional group of the ammonium cation between metal halide

octahedral sheets can be tailored to accommodate neutral molecules. Early work demonstrated flouryl-aryl bonds to be strong enough to form a stable intercalated $(C_6H_5C_2H_4NH_3)_2SnI_4$ compound with hexafluorobenzene between the aliphatic layers (Fig. 16a).[169] Alternatively, the perfluorinated analog ($(C_6F_5C_2F_4NH_3)_2SnI_4$) will uptake benzene molecules to produce a stable compound at room temperature. Other combinations between fluorinated or unfluorinated spacer molecules or intercalants did not form stable compounds, demonstrating the unique nature of the fluoryl-aryl bond. Each intercalated compound shows a small reversible blueshift with respect to the parent perovskitoid spectrum, from 608 nm (2.04 eV) to 595 nm (2.08 eV) (Fig. 16b). The spectral shift was attributed to subtle structural changes induced in the tin(II) iodide sheets by the intercalated molecules.[169]

More pronounced shifts in optical absorption are achieved when the spacer molecule is tuned to uptake more polarizable molecules. Smith *et al.* showed $(IC_6H_{12}NH_3)_2[PbI_4]$, where the spacer molecule contains terminal iodides, will ready uptake $I_2$ to form $(IC_6H_{12}NH_3)_2[PbI_4]\cdot 2I_2$ (Fig. 16c).[170] The authors suggest that the polarizability of $I_2$ molecules decreases the exciton confinement between lead halide layers, resulting in reversible optical absorbance shift of nearly 20 nm (Fig. 16d).

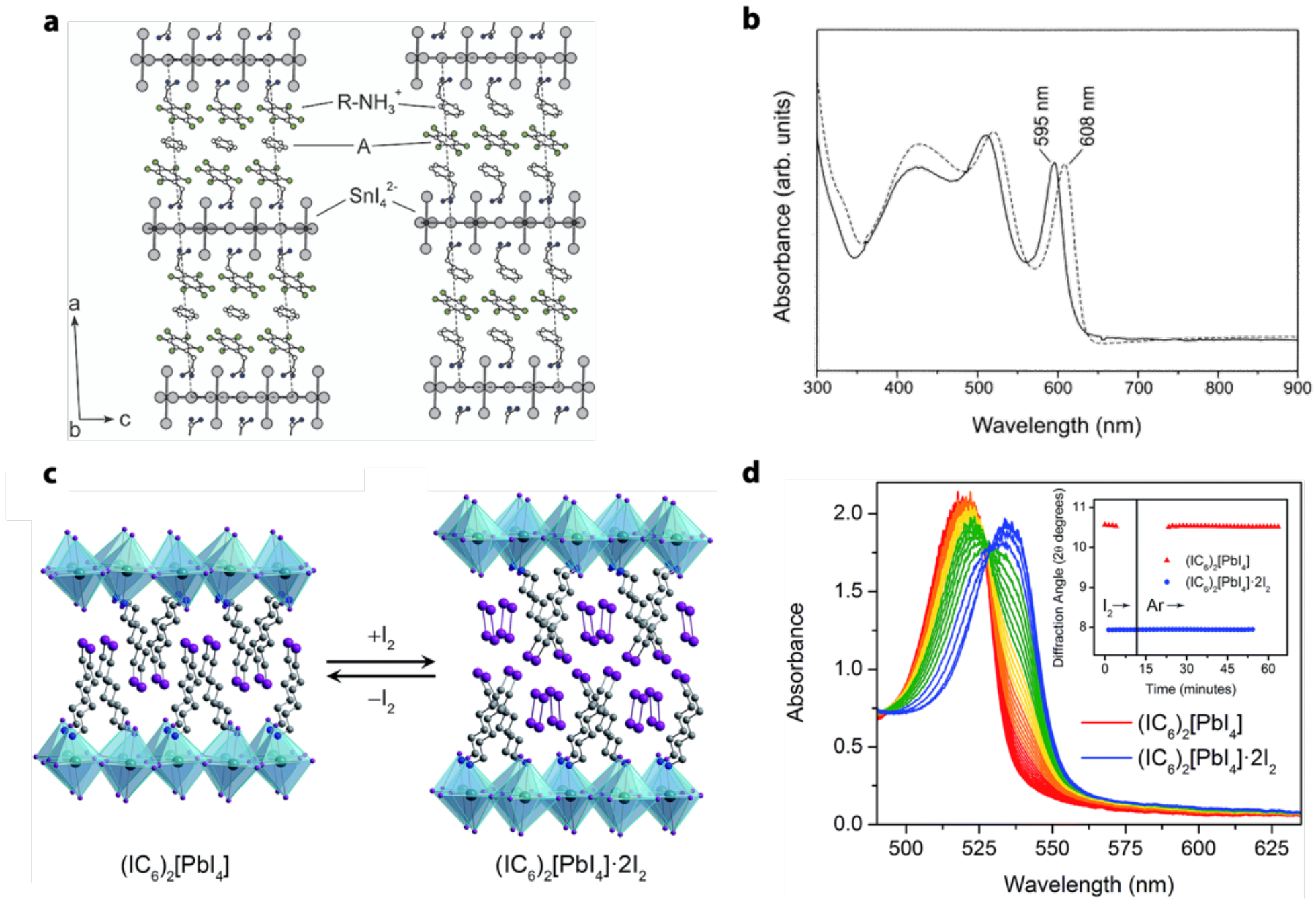

**Figure 16:** (a) Crystal structure of $(C_6H_5C_2H_4NH_3)_2SnI_4$ after intercalation with $C_6F_6$. (b) Absorbance spectra of $(C_6H_5C_2H_4NH_3)_2SnI_4$ before and after intercalation with $C_6F_6$. Adapted from [169] (c) Crystal structure of $(IC_6H_{12}NH_3)_2[PbI_4]$ (denoted as $(IC_6)_2[PbI_4]$) reversibly intercalating $I_2$ molecules. (d) Resulting absorbance spectra as a function of $I_2$ intercalation. Adapted from [170].

Intercalation driven by interactions with the spacer ligand yields small change in optical properties compared to intercalation of molecules that coordinate with the lead halide sublattice. Methylamine ($MA^0$) will readily intercalate into $MAPbI_3$ to form a variety of structures that dramatically disrupt the octahedral network.[171] The result is an amorphous or liquid substance and a variety of solid-state compounds, including: $(MA^0)_4PbI_2$, $[Pb(MA^0)_6]I_2$, $(MA)_5(MA^0)_2Pb_2I_9$ (Fig. 17a). In each compound, the lead halide octahedra characteristic to MHPs no longer exist, as iodide ligands are replaced by methylamine as L-type ligands on the Pb center.

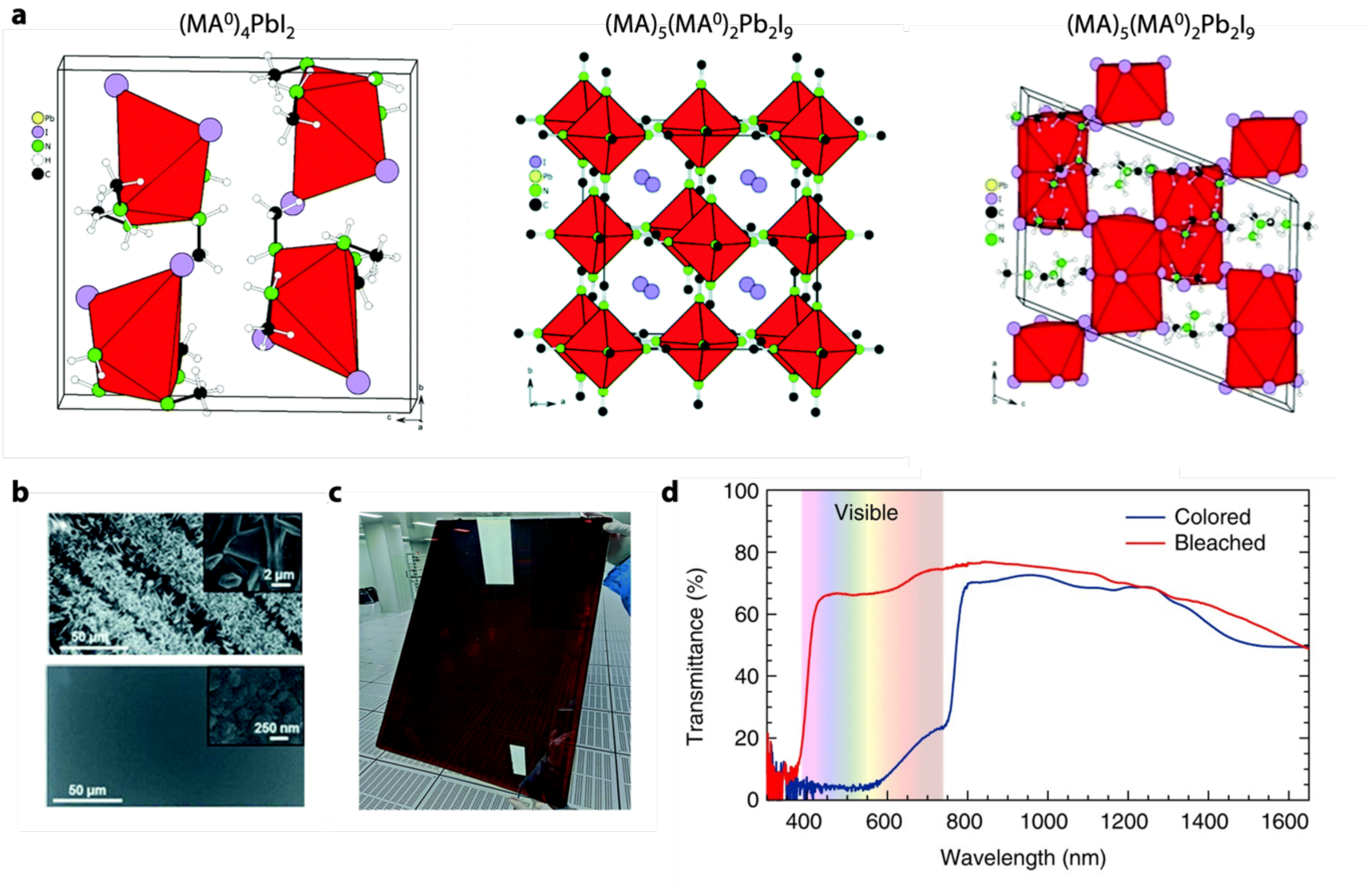


**Figure 17:** (a) Crystal structure of $(MA^0)_4PbI_2$, $[Pb(MA^0)_6]I_2$, and $(MA)_5(MA^0)_2Pb_2I_9$ in (010) direction. $MAPbI_3$ films before (top image) and after (bottom image) methylamine treatment. (c) Photograph of large-area $MAPbI_3$ film. Adapted from [171]. (d) The transformation yields a significant change in the optical transmission of the switchable photovoltaic device. Adapted from [26].

Methylamine has been widely investigated for its ability to liquify and "heal" $MAPbI_3$ films by reducing roughness and growing grain size (Fig. 17b).[172] The process has been shown to

be effective over large areas (Fig. 17c). Intercalation that leads to significant bleaching of optical properties also led to the first report of switchable photovoltaic window technology, which leveraged the effect to achieve a significant change in optical density across visible wavelengths (Fig. 17d).[26] Ammonia and other longer-chain amines have also been shown to reversibly intercalate and bleach $MAPbI_3$.[173][174] The reaction presumably yields similar Pb coordination to the methylamine system, but ammonia and longer-chain amines will freely exchange with methylammonium, which limits the reversibility of $MAPbI_3$ reformation.

### *6.3.2 Ion intercalation*

Extrinsic ion migration has already been discussed in the context of filament formation and altering charge-carrier properties. In this section, we will highlight intrinsic and extrinsic ion motion that leads to significant changes in the optical absorption of the MHP or tunable emission properties for optical write/read applications.

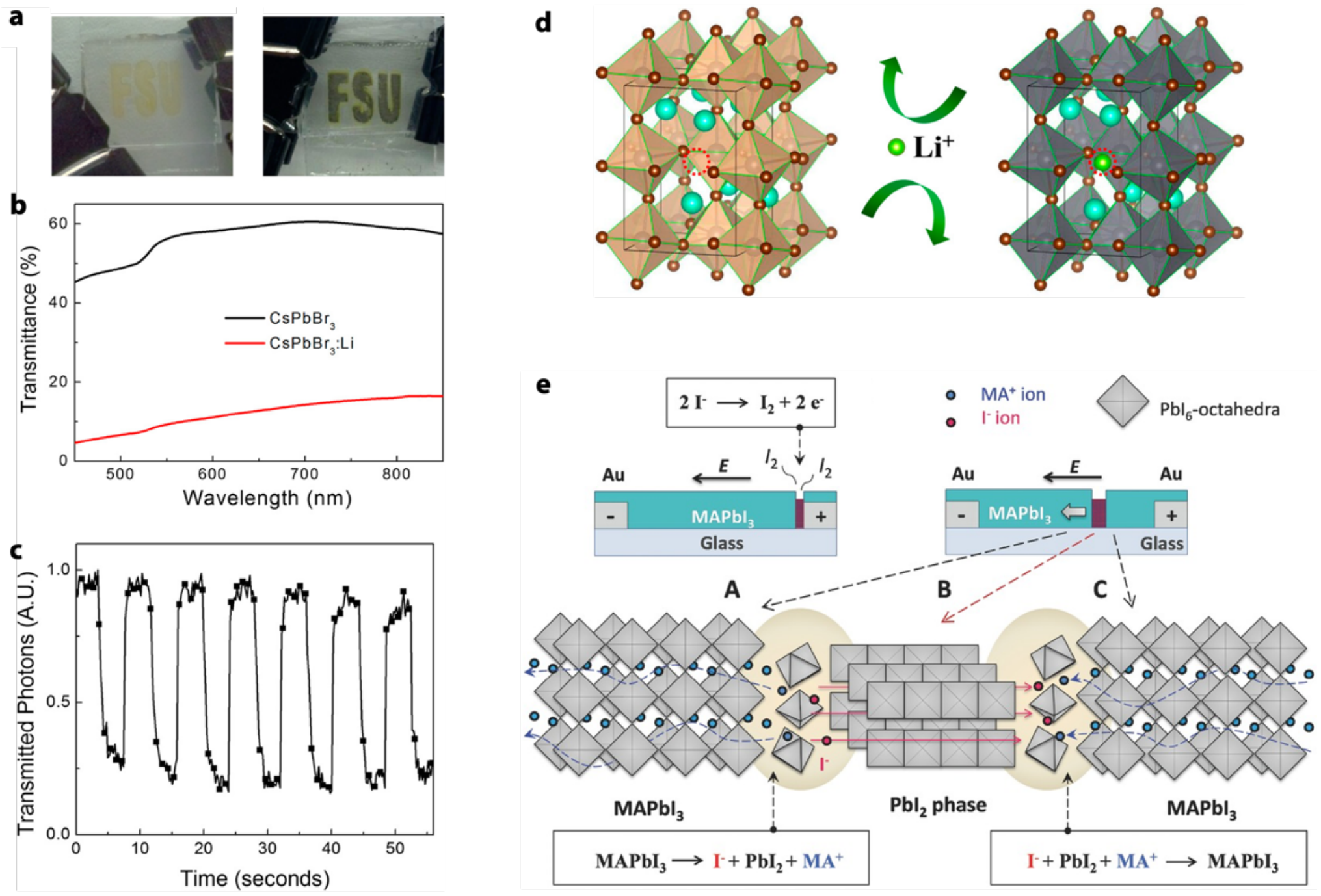


**Figure 18:** (a) Photograph of $CsPbBr_3$ film before and after lithium ion intercalation. (b) Corresponding transmittance spectra of $CsPbBr_3$ films. (c) Transmission over multiple ion intercalation cycles. (d) Schematic of lithium ion intercalation into $CsPbBr_3$. Adapted from Ref. [175] (d) Schematic of electric-field-driven formation of $PbI_2$ due to Iodide and methylammonium ion migration from $MAPbI_3$ Adapted from Ref. [176].

Most examples of intercalation in MHP materials leave the redox state of the MHP constituents unchanged after intercalation. However, the most commonly studied form of intercalation in most chemical system is electrochemical ion intercalation into host materials, as it is the foundation for electrochemical energy storage.[177] There has been initial work to use MHP materials for electrochemical energy storage in lithium ion batteries.[178] Lithiation of $MAPbBr_3$ followed three stages:[179] (1) a small amount of Li ions moves into the halide perovskite lattice by an insertion mechanism, (2) a pure conversion mechanism produces traces of metallic $Pb^0$, and (3) a Li−Pb alloy is formed. Stages 1 and 2 are reversible, whereas stage 3 is an irreversible process. Early work demonstrated this phenomenon in a transparent package, where the metallic $Pb^0$ results in a significant change in the optical density of the film (Fig. 18a,b). The process is repeatable over a number of cycles (Fig. 18c,d).[175]

Reversible ion exchange in MHP materials typically requires two chemical stimuli. However, recent work has shown that mixed halide MHP films composed of iodide and bromide will phase separate iodide-rich and bromide-rich regions upon light illumination[180] or electric field application.[181] The effect produces distinct photo- or electro-luminescence signals that could be applied for memory applications. Since the effect occurs below the threshold of carrier injection, it was concluded that electric field is enough to break the Pb-halide bonds. Electric-field-driven phase transformations have been observed at elevated temperatures, where iodide anions from the $MAPbI_3$ lattice migrate to the cathode, and methylammonium cations migrate to the anode. The area vacated by the ions is macroscopic and yields an optically distinct $PbI_2$ region sandwiched by $MAPbI_3$ that is rich is iodide or methylammonium (Fig. 18e).[176] The effect is reversed by removing the bias or applying the opposite bias, though electrochemical reactions at the electrode interfaces may eventually reduce cycle stability.

#### *6.3.3 Salt pair intercalation*

Salt pairs will readily add to or subtract from the existing metal halide octahedra lattice of MHPs to tune the dimensionality and connectivity of the structures. The phenomenon nicely follows the principle of dimensional reduction, where material stability is predicted based on topochemical addition or removal of salt pairs to metal-anion frameworks.[182] There are many examples of solution-based chemistry where salt pairs intercalate into a $PbX_2$ lattice to produce 3D MHP materials[183] or higher order 0D materials such as $Cs_4PbBr_6$.[146]

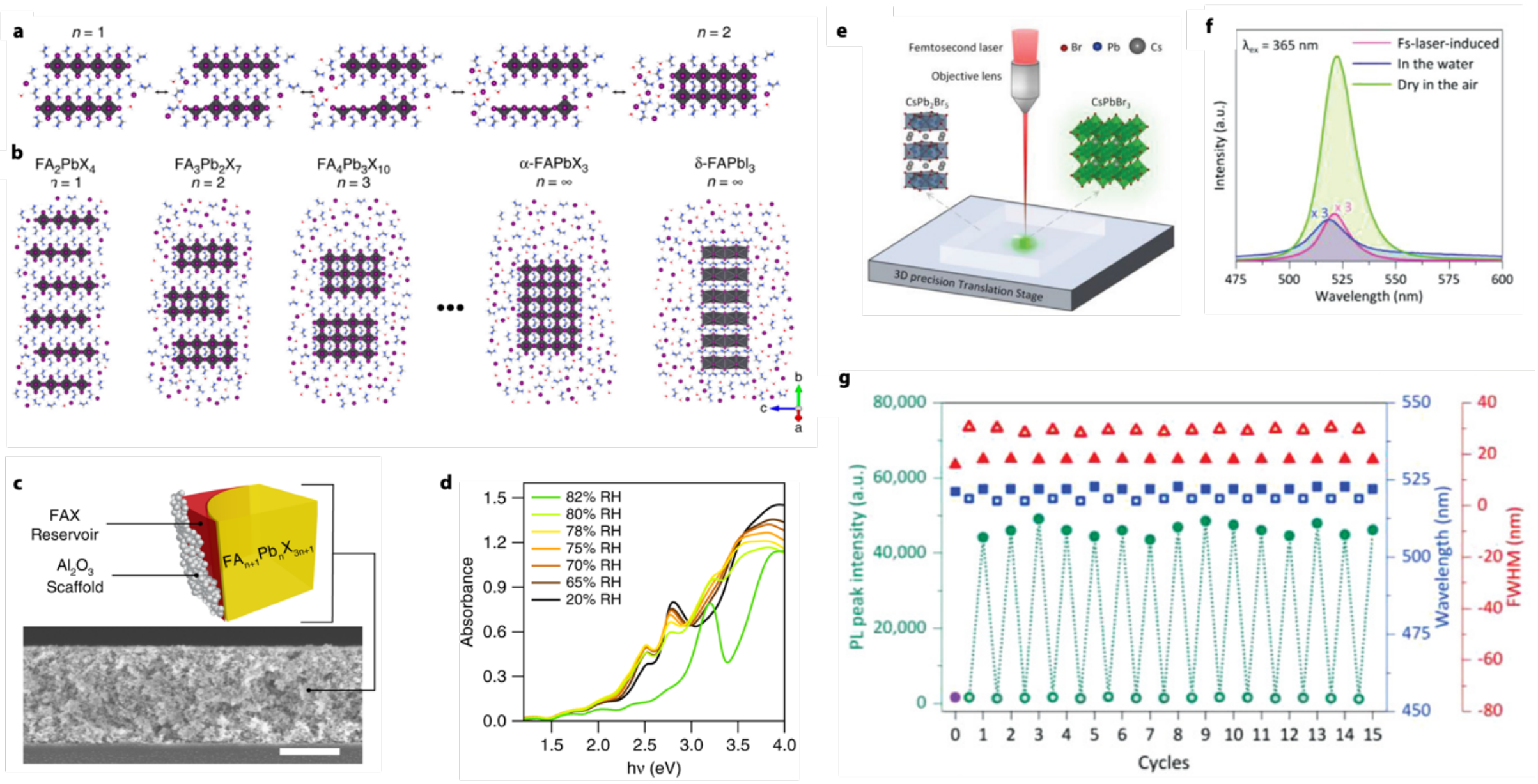


**Figure 19:** (a) Coalescence of two n=1 layers to form an n=2 layer. (b) Schematic of scaffold strategy for enhanced cycle stability. (c) Absorbance spectra of $FA_{n+1}Pb_nI_{3n+1}$ in different layered states controlled by the relative humidity, which induces salt pair intercalation. (d) Dimensionality increase in $FA_{n+1}Pb_nI_{3n+1}$ from n=1 to n= 3 and finally a crystal phase transformation from α-$FAPI_3$ to δ-$FAPbI_3$. Adapted from [108]. (e) Schematic of femtosecond laser-writing of $CsPb_2Br_5$ films to produce $CsPbBr_3$ phases. (f) Photoluminescence of the $CsPb_2Br_5$/$CsPbBr_3$ film before and after laser irradiation. (g) Photoluminescence (PL) peak intensity, Peak wavelength, and full width at half maximum (FWHM) over fifteen write/erase cycles. Adapted from Ref. [184]

A recent example of solid-state dimensional reduction chemistry is the transformation of $FA_{n+1}Pb_nX_{3n+1}$ (FA = formamidinium, X = I, Br; $n$ = number of layers = 1,2,3… ∞) materials in response to humidity or alcohol environment.[108] FAX molecules are removed from a single-layer (n=1) 2D $FA_2PbX_4$ until a thicker n=2 perovskitoid ($FA_3Pb_2X_7$) is produced (Fig. 19a). The mechanism continues to evolve to higher n layers until 3D α-phase $FAPbX_3$ (n=∞) is synthesized (Fig. 19b). When the α-phase $FAPbX_3$ domains reach a critical size, the thermodynamically favored δ-phase will result. The effect is reversed with heat. Durability to

repeated cycling of the optical states is enhanced with an optically inactive scaffold (Figure 19c), which houses $FA_{n+1}Pb_nX_{3n+1}$ as well as a "reservoir" of FAX molecules that are intercalated and deintercalated from the $FA_{n+1}Pb_nX_{3n+1}$ domains. The optical properties of $FA_{n+1}Pb_nX_{3n+1}$ films exhibit multicolor chromism, where films contain multiple different n layers that evolve with varied humidity or alcohol environments to produce a near-continuum of optically distinct states that could be leveraged for memory or optical read/write schemes (Fig. 19d). A closed system can be produced where water or alcohol vapor is trapped within the volume above the film. In the closed system, only heating or cooling stimuli need to be applied to achieve the salt pair intercalation mechanism. A similar strategy was recently employed in the $FAPbBr_3$ system to realize reversible photochromic sensors capable of self-adaptation to lighting.[185]

There are many examples of reversible reconfiguration of the Cs-Pb-Br system to tune emissive and absorption properties. Exposing a $Cs_4PbBr_6$ film to water will precipitate $CsPbBr_3$ nanocrystals to transform the film from dark to highly luminescent.[186][187] Nanocrystals are produced by de-intercalating three equivalents of CsBr salt pairs from $Cs_4PbBr_6$ regions of the film. Exposing a composite film of $CsPbBr_3$ nanocrytals embedded in $Cs_4PbBr_6$ system to water quenches emission by precipitating a higher density of $CsPbBr_3$ nanocrystals that couple together or have reduced surface passivation, which serves as an alternative mechanism for emissive control.[188] A similar transformation occurs when exposing 3D perovskite $CsPbBr_3$ films with excess $PbBr_2$ to water to produce stable 2D perovskitoid $CsPb_2Br_5$.[186] Water initiates intercalation of $PbBr_2$ into the $CsPbBr_3$ lattice. The transformation is accompanied by enhanced absorbance in the visible spectrum and bright photoluminescence that can be tuned by nearly 100 nm using anion alloying with Cl. The $PbBr_2 + CsPbBr_3 \leftrightharpoons CsPb_2Br_5$ transformation has been applied to direct writing of $CsPb_2Br_5$ patterns into $CsPbBr_3$ using femtosecond laser irradiation (Fig. 19e).[184] As shown in previous studies, $CsPb_2Br_5$ absorption

is redshifted and exhibits stronger photoluminescence compared to $CsPbBr_3$ (Fig. 19f). The transformation is reversed by exposing the film to water, which can be repeated over many cycles (Fig. 19g). The authors extended the optical storage mechanism to three and four dimensions by realizing spatially and temporally resolved optical encryption.

*6.3.4 Co-intercalation of solvated salt pairs*

We distinguish co-intercalation from other mechanisms because neutral coordinating molecules are (de)intercalated along with salt pairs into a 3D perovskite or 2D perovskitoid to produce thermodynamically stable structures with (lower) higher dimensionality. Co-intercalation is perhaps the first switching mechanism reported in MHP literature. The dihydrate of methylammonium lead iodide single crystal was reported in 1987, well before the excitement around $MAPbI_3$ as a photovoltaic material. The original paper reports dehydration of $MA_4PbI_6{\bullet}2H_2O$ to yield $MAPbI_3$ crystals with distinct optical properties (Fig. 20a).[189] In order to transform $MAPbX_3$ into the dihydrate, $3MAX{\bullet}2H_2O$ must be added to the crystal. When discussed in the photovoltaics literature, it is often written as a MAX-deficient reaction, which produces $PbI_2$ in the film to emphasize the effect of water on mechanisms of degradation (e.g. $MAPbI_3 \leftrightharpoons MA_4PbI_6{\bullet}2H_2O + PbI_2$).[190]

To capitalize on co-intercalation for memory or repeatable switching, excess AX salt is added to MHP films, and an oxide scaffold is often used to improve switching durability. In this case, the reversible reaction is: $MAPbX_3 + 3MAX{\bullet}2H_2O \leftrightharpoons MA_4PX_6{\bullet}2H_2O$. Halder et al. first demonstrated reversible reaction for thermochromism in $MAPbI_3$ solar cells,[191] and there have been many reports since with similar chemistries and application to thermochromic windows.[81,83,192–194] Reversible co-intercalation of hydrated MAI was also recently shown to occur in layered 2D analogs of $MAPbI_3$ and show robust cycling behavior.[193] Similar to methylamine demonstrations, water vapor can be sealed into a vessel with the perovskite composite film to produce a closed system that may be switched with heat as the sole stimulus.

Interestingly, there have been reports that after a MAX-rich $MAPbX_3$ film has been converted to $MA_4PbX_6•2H_2O$ by increasing relative humidity to >60%, the film will then respond to further increase in relative humidity by de-intercalating $3MAX•2H_2O$ to produce $MAPbX_3$ crystals in the film.[195] The result suggests a complex relationship between the thermodynamically favored phase and the impact of hydrogen bonding environments.

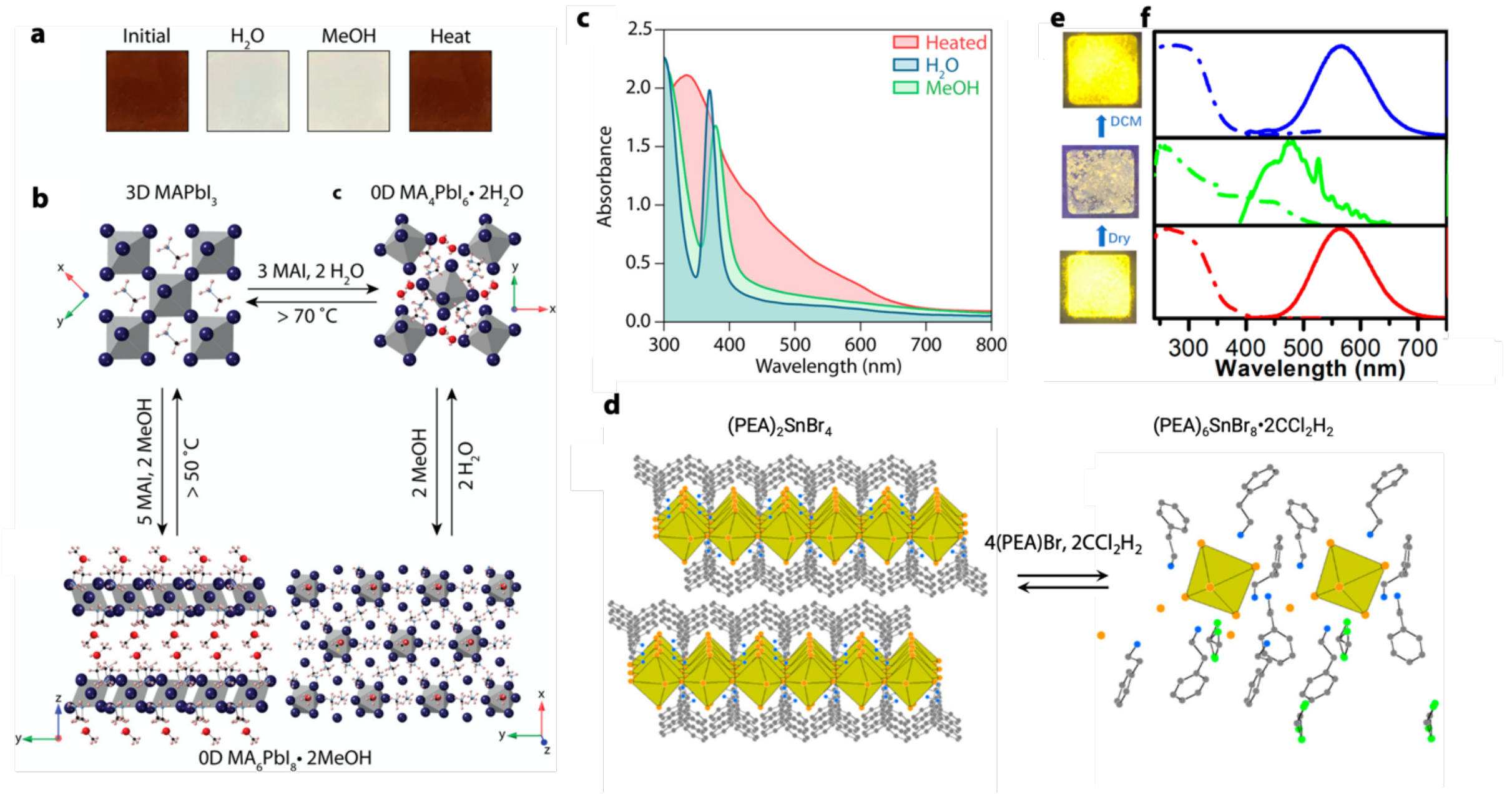


**Figure 20:** (a) Images of a $MAPI_3$ composite film before exposure to $H_2O$ or MeOH to produce bleached films by co-intercalation of MAI and $H_2O$ or MeOH. Heat returns the film to its original state. (b) Illustration comparing the crystal structures of 3D $MAPbI_3$, 0D $MA_4PbI_6 \cdot 2H_2O$ hydrate complex, and 0D $MA_6PbI_8 \cdot 2MeOH$ methanolate complex[28]. Absorbance spectra of 3D $MAPbI_3$, 0D $MA_4PbI_6 \cdot 2H_2O$ hydrate complex, and 0D $MA_6PbI_8 \cdot 2MeOH$ methanolate complex. (d) Illustration of the reversible $(PEA)_2SnBr_4•2CCl_2H_2$ to $(PEA)_6SnBr_8•2CCl_2H_2$ transformation due to co-intercalation of 4(PEA)Br, $2CCl_2H_2$. (e-f) Images (e) and normalized absorptance and emission (f) before and after intercalation of (PEA)Br and $CCl_2H_2$. Adapted from [196].

There are new reports of alternative co-intercalation chemistries to the hydrate system. Analogous to hydrate formation, methanol was recently shown to form a stable structure of the form $MA_6PbX_8•2MeOH$ (Fig. 20b).[28] Methanol is intercalated with MAI salt pairs to produce a 0D perovskite where lead iodide octahedra are spatially isolated in planes of the crystal instead of isotropically like the dihydrate materials (Fig. 19b). The methanolated material exhibits an absorption band in the UV, which is slightly redshifted compared to the hydrate analog (Fig. 20c). The relative crystallographic positions of the metal halide octahedra in the

structure of $MA_6PbX_8$•2MeOH is the same as $(PEA)_6SnBr_8$•$2CCl_2H_2$ (Fig. 19d).[196] $(PEA)_6SnBr_8$•$2CCl_2H_2$ films exhibit yellow emission and a large Stoke's shift from the UV absorption onset. De-intercalation of 4(PEA)Br•$2CCl_2H_2$ produces a film with a blue-shifted emission and redshifted absorption to significantly reduce the Stoke's shift of the intercalated material (Fig. 20 e,f).

## 7. Future Outlook and Perspective

### *7.1 Correctly Accounting for, and Optimizing, Energy Efficiency*

Since the ultimate goal of neuromorphic architectures is to process information at a fraction of the energy consumption of traditional CMOS paradigms, it is useful to consider the energy consumption of optically stimulated neuromorphic elements. It is first important to consider whether the device is intended to serve as a neuromorphic sensor, to perform computations, or both. In platforms intended for sensing of external optical stimuli, the photons delivered to the device can largely be considered as 'free' stimuli, since they do not need to be delivered by a controlled source such as an LED or laser. In platforms intended for computing, where highly controlled pulse sequences are used to program and/or manipulate synaptic weights, the energy consumption of the photon delivery source must be accounted for.

The energy consumption ($\Delta E$) of optically stimulated devices is often calculated by considering the optical pulse energy delivered directly to a device:

$$E = P \times \Delta t \times A_d \qquad \text{Equation 2}$$

where $P$ and $\Delta t$ are the power density (W/cm$^2$) and temporal duration (s) of the incident light pulse and $A_d$ is the area of the device (cm$^2$). It should be noted however that this is the *lower limit of* $\Delta E$, since it does not consider any photons outside of the device area that are not absorbed and converted to the photocurrent. If the photons are not 'free' and must be delivered in a controlled fashion *via* LED or laser pulses, *all* photons delivered in a pulse must be counted towards energy consumption by replacing device area ($A_d$) with pulse area ($A_p$). Ultimately, an

accurate accounting would actually account for the inefficiencies of the pulse source by also considering the total electrical energy consumed by the illumination source.

The electrical energy consumption of electrically read neuromorphic elements *during the stimulus pulse* is often calculated by:

$$\Delta \mathrm{E} = \mathrm{V} \times I \times \Delta t \qquad \text{Equation 3}$$

where $V$ and $I$ are the voltage and current of the device, respectively, and $\Delta t$ is the stimulus pulse duration. However, it is important to note that some three-terminal devices require the application of substantial (often constantly applied) gate voltages, which Equation 3 does not account for. Ultimately, for optically stimulated and electrically read elements, both the photonic and electrical energy consumption should be accounted for to provide an accurate comparison to traditional CMOS approaches. Therefore, the following points need to be considered: (1) device minimization for densely packed elements, (2) waveguiding and/or other optical engineering strategies to maximize the delivery efficiency of photons to small synaptic elements, and (3) energy-efficient pulsed sources, and (4) low operating voltages (both source/drain *and* gate voltages, if three-terminal).

**Table 1. Summary of Reported Metrics for Optical Memory, Switching, and Neuromorphic Devices**

| | **Active Layer Material** | **Dimensions (D or W × L) (µm²)** | **Area (µm²)** | **λ (nm)** | **Optical Pulse Intensity (µW cm⁻²)** | **Optical Pulse Width (ms)** | **Energy consumption (µJ)** | **Ref.** |
|---|---|---|---|---|---|---|---|---|
| **Metal Halide Perovskites** | $(PEA)_2SnI_4$ | 1800 x 800 | 1440000 | 470 | 39.2 | 200 | 1.129E-01 | [100] |
| | $(PEA)_2SnI_4$ | 1800 x 800 | 1440000 | 470 | 39.2 | 10 | 5.645E-03 | [100] |
| | $(PEA)_2SnI_4$ | 1800 x 800 | 1440000 | 470 | 3.4 | 10 | 4.896E-04 | [100] |
| | $(PEA)_2SnI_4$ | 1800 x 800 | 1440000 | 470 | 39.2 | 1 | 5.645E-04 | [100] |
| | $CsPbBr_3$ | 50 x 1000 | 50000 | 365 | 41 | 1000 | 2.050E-02 | [61] |
| | $MAPbI_3$ | Diameter of Ag 100 µm | 7854 | - | 3750 | 400 | 1.178E-01 | [33] |
| | $CsPbBr_3$ | Diameter of Ag 100 µm | 7854 | - | 8160 | 300 | 1.923E-01 | [33] |
| | $CsPbBr_3$ | 30 x 1000 | 30000 | 500 | 100 | 50 | 1.500E-03 | [197] |
| | $(PEA)_2SnI_4$ | 7 x 50 | 350 | 470 | 5 | 20 | 3.500E-07 | [198] |
| | $CsPbBr_3$ | 7 x 50 | 350 | 470 | 11.6 | 20 | 8.120E-07 | [198] |
| | $(PEA)_2SnI_4$ | 7 x 50 | 350 | 470 | 5 | 20 | 3.500E-07 | [198] |
| | $MAPbI_3$ | 25 × 500 | 12500 | 532 | 1 | 200 | 2.500E-05 | [130] |
| | $Cs_2Pb(SCN)_2Br_2$ | 50 x 1000 | 50000 | 450 | 180 | 60000 | 5.400E+00 | [199] |
| | $Cs_2Pb(SCN)_2Br_2$ | 50 x 1000 | 50000 | 450 | 180 | 60000 | 5.400E+00 | [199] |
| | $CsPbBr_3$ | 30 x 1000 | 30000 | 450 | 260 | 50 | 3.900E-03 | [128] |
| | $CsPbBr_3$ | 30 x 1000 | 30000 | 450 | 50 | 50 | 7.500E-04 | [128] |
| | $FAPbI_3$ | 5 x 1000 | 5000 | 405 | 50 | 0.03 | 7.500E-08 | [75] |
| | $FAPbI_3$ | 0.1 x 1000 | 0.1 | 405 | 7400 | 1 | 7.400E-09 | [75] |
| | $Cs_2AgBiBr_6$ | 300 x 100 | 30000 | 365 | 12700 | 30000 | 1.143E+02 | [134] |
| | $Cs_2AgBiBr_6$ | 300 x 100 | 30000 | 455 | 12700 | 30000 | 1.143E+02 | [134] |
| | $Cs_2AgBiBr_6$ | 300 x 100 | 30000 | 505 | 12700 | 30000 | 1.143E+02 | [134] |
| | $Cs_2AgBiBr_6$ | 300 x 100 | 30000 | 660 | 12700 | 30000 | 1.143E+02 | [134] |
| | $Cs_2AgBiBr_6$ | 300 x 100 | 30000 | 455 | 12700 | 1000 | 3.810E+00 | [134] |
| | $Cs_2AgBiBr_6$ | 300 x 100 | 30000 | 455 | 3200 | 30000 | 2.880E+01 | [134] |
| **Metal Oxides** | IGZO | 10 x 10 | 100 | 405 | 7230 | 50 | 3.615E-04 | [200] |
| | IGZO | 800 × 500 | 400000 | 450 | 15 | 500 | 3.000E-02 | [201] |
| | IGZO | 100 x 20 | 2000 | 465, 525, 620 | 22 | 50 | 2.200E-05 | [202] |
| | IGZO | 10 x100 | 1000 | 254 | 200 | 1000 | 2.000E-03 | [203] |
| | $IGZO/Al_2O_3$ | 20 x40 | 800 | 365 | 3000 | 100 | 2.400E-03 | [204] |
| | IGZO, ISZO, ISO, IZO | 180 x 70 (100 x 10) | 1000 | 380-385 | 600 | 500 | 3.000E-03 | [205] |
| | IGZO | 80 x1600 | 128000 | 275, 365, 405 | 18.23 | 100 | 2.333E-03 | [206] |
| | IGZO/chitosan | 80 x1000 | 80000 | 405 | 176000 | 20 | 2.816E+00 | [207] |
| | ITO/chitosan | 80 x1000 | 80000 | 405 | 45 | 500 | 1.800E-02 | [208] |
| | $MoO_x$ | Diameter of 250 um | 49087 | 365 | 220 | 200 | 2.160E-02 | [34] |
| | $In_2O_3$ | 80 x1600 | 128000 | 365 | 910 | 80 | 9.318E-02 | [209] |
| | $CeO_{2-x}$ | Diameter of 100 µm | 7853 | 600 | 60 | 4000 | 1.885E-02 | [210] |
| | $CeO_{2-x}$ | 200 µm | 31415 | 499, 560, 638 | 0.3 | 10000 | 9.425E-04 | [211] |
| | $ZnO_{1-x}$ | Diameter of 100 µm | 7854 | 310 | 72 | 1000 | 5.655E-03 | [212] |
| | ZnO | 6 x $10^7$ µm$^2$ | 60000000 | 365 | 1000 | 1000 | 6.00E-04 | [213] |
| | $La_{1.875}Sr_{0.125}NiO_4$ | 90000 µm$^2$ | 90000 | 365, 450, 520, 730 | 20000 | 5000 | 9.000E+01 | [214] |
| | Nb: $SrTiO_3$ | 100 um | 7854 | 459, 528, 630 | 10000000 | 500 | 3.927E+02 | [215] |
| | $BiFeO_3/SrRuO_3$ | 25 x 25 | 625 | 405, 532, 1064 | 10000 | 60000 | 3.750E+00 | [216] |
| **Metal Chalcogenides** | $MoS_2$ | 15 × 10 | 150 | 473, 532, 655 | 50000 | 100 | 7.500E-03 | [217] |
| | $MoS_2$ | 9 x 20 | 180 | 445 | 52.5 | 10000 | 9.450E-04 | [36] |
| | $MoS_2$ | Diameter of 50 µm | 1963.5 | 310 | 110 | 1000 | 2.160E-03 | [218] |
| | CdS | $2.5 \times 10^4$ | 25000 | 365, 465, 525 | 100000 | 500 | 1.250E+01 | [219] |
| | $MoS_2$/ PTCDA | 2 x 5.3 | 10.6 | 532 | 127000 | 10000 | 1.346E-01 | [220] |
| **Organic Semiconductor** | C8-BTBT | 40 x 20 | 800 | 365 | 9 | 50 | 3.600E-06 | [221] |
| | DPPDTT/ $CsPbBr_3$ | 30 x 1000 | 30000 | 450 | 260 | 1000 | 7.800E-02 | [128] |
| | Sol-PDI, BPE-PDI | 50 x 1000 | 50000 | 405, 530, 640 | 10000 | 3000 | 1.500E+01 | [222] |
| | C8-BTBT/ PAN | 200 x 6000 | 1200000 | 360 | 900 | 200 | 2.160E+00 | [223] |
| | BP | 9.2 ± 1.1 | 9.2 | 280 | 400 | 100 | 3.680E-06 | [224] |
| | BP-ZnO hybrid NPs/ pentacene | 50 x1000 | 50000 | 365/ 520/ 660 | 150000 | 1000 | 7.500E+01 | [225] |
| **Nanocarbon** | Graphene/ SWNTs | 30 x 90 | 2700 | 405/ 532 | 70735 | 5 | 9.549E-03 | [54] |
| **Silicon Nanocrystals** | Si NCs | 2000 x 2000 | 4000000 | 375, 532, 1342, 1870 | 18000 | 2000 | 1.440E+03 | [226] |
| | Si NCs | 10 x120 | 1200 | 375, 532, 1342 | 1300 | 200 | 3.120E-03 | [56] |

Table 1 and Figure 21 summarize the current state of research on optically stimulated synaptic devices based on both MHPs and several other classes of materials and heterojunctions. The optical energy consumption in Figure 21a is calculated with Equation 2 and normalized by device area. Figure 21a demonstrates that MHP-based devices have shown some of the lowest optical energy consumption values *and* have been operated with some of the shortest optical pulse widths across all currently studied material classes. These low values bode well for MHP-based optical synapses, since two important goals for optical neuromorphic applications are low-energy operation and fast switching times. Furthermore, since MHPs can respond to much shorter optical pulses than have been demonstrated so far for optical synaptic devices, there is the potential to realize fast switching times (sub-nanosecond) that surpass those possible with electronic stimuli or indeed in their biological counterparts. Figure 21b illustrates that synaptic devices typically respond to optical stimuli in the visible portion of the electromagnetic spectrum (ca. 380 – 700 nm), with only a few example of material systems that exhibit a response at longer wavelengths. Figure 21b demonstrates a clear need for exploring MHP-based devices that respond to longer wavelengths in the visible and/or near-infrared regions of the spectrum, since the majority of studies have thus-far demonstrated devices operating at relatively short wavelengths. In this regard, Figure 3a-b illustrates the potential that MHPs possess for tunable optical switching behavior, with the ability to alter the optical absorption via modification of the composition and dimensionality. For example, there are a growing number of demonstrations of efficient opto-electronic devices that utilize low bandgap MHPs (e.g., mixed Sn-Pb hybrid organic-inorganic perovskites) with absorbance and emission in the near-IR. For additional reading on power consumption efficiencies of photon-based information processing and neuromorphic architectures, we point the reader to several recent analyses and reviews.[227–229]

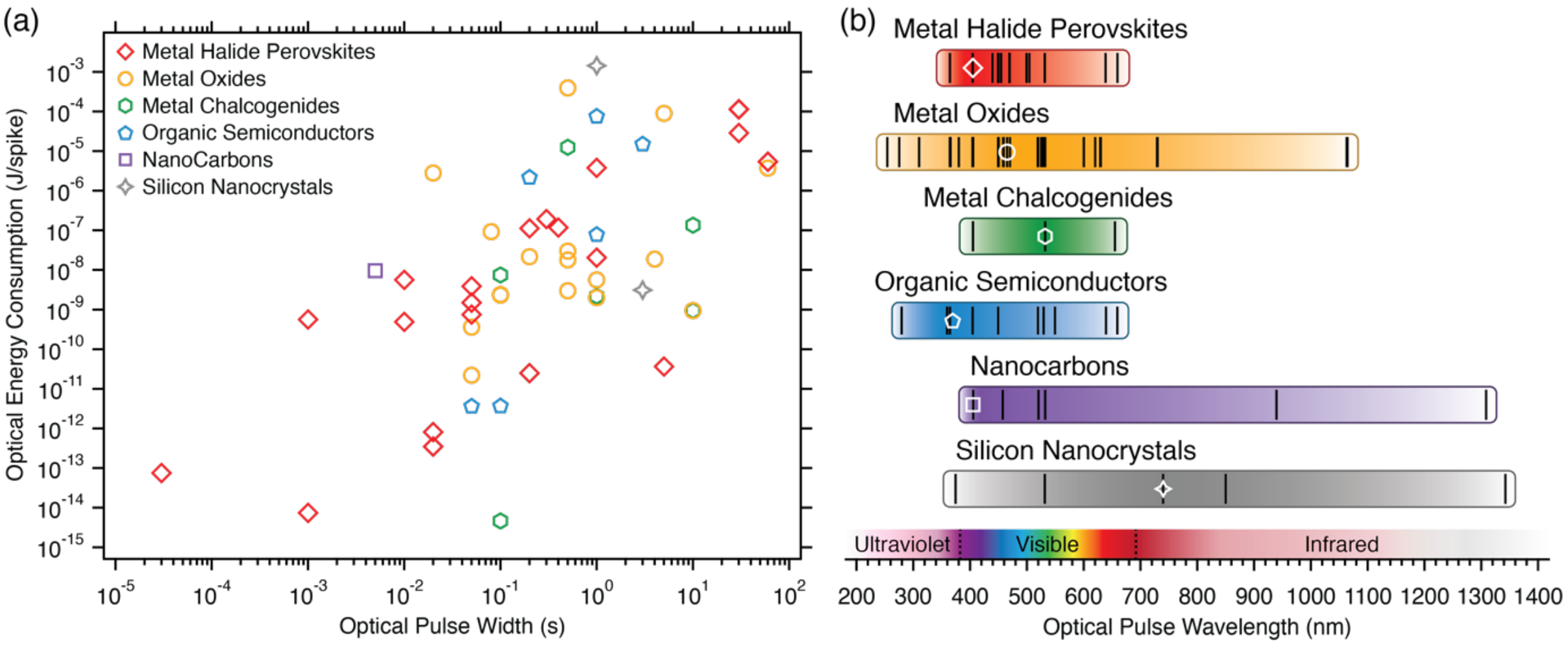

**Figure 21. (a)** Summary of optical energy consumption (J/spike), as a function of the optical pulse width used to stimulate the device, for optically stimulated synapse-like devices constructed from a broad variety of materials/heterojunctions. **(b)** Comparison of the wavelength range used to stimulate the devices summarized in panel (a). The colored bar represents the full wavelength range of the optical stimuli employed in these materials. The black vertical lines illustrate the wavelengths employed in the individual studies and the symbols illustrate the wavelength employed for the device in each material class

that exhibits the lowest measured optical energy consumption. All of the individual metrics used to generate these panels are also summarized in Table 1.

### *7.2 Understanding and Utilizing New Mechanisms*

Despite some early successes for MHP-based artificial synapses, key knowledge gaps exist, including: (i) the roles of light, MHP dimensionality, and key interfaces in controlling and modulating the dynamics of charge trapping and ion migration; (ii) the precise ion/charge transport mechanisms and atomistic pathways; (iii) strategies for repeatable, minimally stochastic R-C modulation and control over temporal dynamics. The field has also barely scratched the surface with regards to the breadth of unique heterostructures that can be employed for MHP-based optical memory devices. Such heterostructures may be between MHPs and other semiconductors, semi-metals, metals, or degenerate semiconductors. One area which may prove promising is employing heterostructures between two or more MHP layers with distinct compositions, dimensionalities, bandgaps, and so on. These types of heterostructures are challenging to synthesize and stabilize, due in large part to facile ion mixing across interfaces, but such interfaces could enable versatile neuromorphic devices.

Beyond improved mechanistic understanding of current implementations, new opportunities exist for MHP-based neuromorphic materials and devices that utilize unique mechanisms. Some reports on $MAPbI_3$ suggest that ion migration results in chemical segregation and ferroelastic domains.[230,231] Though several MHPs are proven to be ferroelectric, ion migration in some non-ferroelectric MHPs is sometimes confused with ferroelectricity.[99] Yuan and co-workers[72] reported that ion accumulation across grain boundaries of polycrystalline MHPs could increase above-bandgap photocurrent due to the formation of randomly dispersed tunnelling junctions across grain boundaries of the polycrystalline films. The resulting photovoltage is analogous to the Anomalous photovoltaic effect[99] which is often reported in ferroelectrics. Such phenomena could be very useful for ultra-fast photonic neuromorphic applications. Therefore, it is interesting to consider what can be achieved by the presence of ferroelectricity or similar phenomena in MHPs.[98,99]

While we have discussed a number of powerful strategies by which changes in the optical properties or emission of MHP absorbers can be tuned and read out as synaptic weight, the incorporation of these mechanisms and devices into memory and neuromorphic architectures is at a very early stage. We note that these structural changes should also dramatically modulate electrical properties, although this line of research is even less mature than the study of optical property switching and represents a potentially fruitful area of future research.

It is also interesting to consider the additional degrees of freedom available to the photons for optically stimulated and/or read MHP-based neuromorphic elements. In Figure 11, we briefly touched on the use of chiral organic A-site cations or ligands for introducing chirality into the MHP lattice. This strategy can enable optically stimulated neuromorphic elements where the handedness of the incident photons provides another sensitive handle for manipulating synaptic weight. For optically read synaptic elements, it is also important to consider that changes in transmission are not the only way to read out synaptic weight. The photon energy, intensity, spatial distribution, optical polarization, and frequency of *emitted* photons could also be used to manipulate and process information. Interesting concepts for memory elements utilizing photon emission have been demonstrated for other material systems, including excitonic switches[232–234] and electroluminescent synapses.[55] MHPs have been incorporated into efficient light-emitting devices[235–237] and chiral MHPs have even been shown to enable significant anisotropies for emitting right- or left-handed photons in e.g. "spin LEDs".[238] Given that optical communication is more energy-efficient and faster than electrical transmission, and is a widely accepted and mature strategy for information processing, bidirectional (absorbing and emitting) neuromorphic devices are in high demand. However, MHPs have yet to be explored in this direction.

While the optical transmission and emission strategies discussed above are naturally centered around the ultraviolet, visible, and near-infrared versions of the electromagnetic spectrum, additional opportunities exist for tuning and reading out synaptic weight of MHP neuromorphic devices in other frequency regimes, such as in the GHz and THz regions. Hao et al. demonstrated that synaptic weight programmed by 4 ns light pulses in three-terminal MHP nanocrystal heterojunction synapses could be read out by a CW microwave (9 GHz) probe.[75] Such microwave probe schemes, potentially in either transmission or reflection geometries, may be scalable for miniaturized devices that can operate at high speeds, since a number of options exist for on-chip microwave resonators. Optical conductivity can also be probed in the THz frequency regime, as has been implemented in a proof-of-concept demonstration for the prototypical $Ge_2SbTe_5$ (GST) phase change material.[239] Promisingly, there have also been demonstration of all-optical switching in hybrid organic-inorganic perovskite semiconductors in the THz frequency domain.[240]

*7.3 Towards Complex Arrays and Neural Networks*

The integration of MHP-based three-terminal neuromorphic devices assisted by light illumination in large scale arrays is still a technological challenge and warrants immediate attention. John et al.[96] provide a compelling recent example of using electrically stimulated synaptic behavior in higher order MHP arrays for reservoir computing by utilising ionotronic diffusion and drift[101] in $CsPbBr_3$ NCs. The device characteristics were used to run a network simulation based on reservoir computing. The study suggested that distinct computing approaches such as reservoir computing, spiking neural networks and artificial neural networks rely on volatile, non-volatile or a combination of two types of components. The authors were able to achieve both volatile diffusive and multi-state non-volatile drift kinetics in the same memristor through the manipulation of compliance currents. Simulations were carried out to demonstrate the advantages of the reconfigurability features using a fully-memristive reservoir computing architecture. The framework used a dynamically-configured layer of virtual volatile memristors (reservoir nodes) and an artificial neural network layer readout with non-volatile weights. The reservoir framework was recorded to achieve training and test accuracies of 86.75% and 85.14% between 2- 5 Epochs, respectively.

The speed-retention dilemma in many mixed ionic/electronic conducting neuromorphic elements dictates that while low ion activation energies are beneficial from the standpoint of switching speed/energy, they could also lead to poor retention of the programmed state because ions can thermally re-equilibrate when the stimulus is removed. Light energy could potentially be used to program arrays of MHP neuromorphic elements in predictable and controllable ways that may help overcome this traditional speed-retention dilemma.

The widespread implementation of neuromorphic computing architectures based on MHPs (and indeed any active material class) is likely to require efforts focused on material scaling and device miniaturization. Miniaturization of MHPs for optical memory and neuromorphic applications is favorable compared to other materials due to the high absorption coefficient, which allows for smaller material volumes. Wheras many of the proof-of-concept MHP-based neuromorphic devices demonstrated thus far are fairly large area, there are already a number examples suggesting that such miniaturization is feasible[241]. For example, MHP-based optical memory devices have been demonstrated via fabrication of short channels (length $\lesssim$300 nm) comprising a $FAPbBr_3$ NC/(6,5) SWCNT heterojunction[75] and MHP active layers have also been incorporated into photovoltaic architectures that have been processed using conventional photolithographic techniques[242] and more scalable techniques[243]. As with any optical technique, miniaturization is ultimately constrained by the diffraction limit in the far field ( $\lambda/2NA$, where NA is the numerical aperture), so that devices with dimensions less than ~λ/2 would require near-field techniques to fully utilize the energy density within incident light pulses.

*7.4 Non-Pb-based Metal Halide Materials*

Pb-based MHPs have seen the most research attention due to their success as photovoltaic absorber materials. Efforts to completely remove Pb have resulted in poorer photovoltaic performance. However, optical memory, switching and neuromorphic functionality may not require the same optoelectronic features that distinguish Pb-based compounds from their alternatives in the PV realm. There are now many examples of optical switching behavior in Cu-,[244–247] Bi-,[248] Sb-halide,[249,250] and double perovskite compounds[251] (such as $Cs_2AgBiBr_6$) that feature two metal centers.[252][253] Each of these metal centers has multiple stable oxidations states, which points to the likelihood of an even richer diversity of switching chemistries compared to the Pb-halide system reviewed here that tend toward $Pb^0$ or $Pb^{2+}$ states. To date, most reversible *structurally-driven* switching behavior is due to crystal phase transformations in lower-dimensional compounds (1D or 2D) where octahedral tilting and bond distortion are the mechanisms of changes in optical state. The connectivity of the metal halide polyhedra remains intact. The other mechanisms outlined here (crystalline phase change that leads to changes in connectivity, crystal-to-amorphous transitions, precipitation, and each type of intercalation) all remain largely unexplored and point to a rich opportunity for plentiful research in these materials specific to optical memory, switching, and neuromorphic behavior.

**Acknowledgements**

This work was authored by the National Renewable Energy Laboratory (NREL), operated by Alliance for Sustainable Energy, LLC, for the U.S. Department of Energy (DOE) under Contract No. DE-AC36-08GO28308. This work was supported by the Laboratory Directed Research and Development (LDRD) Program at NREL. The views expressed in the article do not necessarily represent the views of the DOE or the U.S. Government. Gaurav Vats acknowledges funding from the European Union's Horizon 2020 research and innovation programme under the Marie Skłodowska-Curie grant agreement No. 892669.